\documentclass[aps,pre,epsf,preprint,showpacs]{revtex4-1}
\usepackage{amsmath}
\usepackage{amssymb}
\usepackage{epsfig}
\usepackage{multirow}
\usepackage{graphicx}
\usepackage{color}

\begin{document}
\title{Monte Carlo study of the two-dimensional kinetic Blume-Capel model in a quenched random crystal field}

\author{Alexandros Vasilopoulos$^1$}
\author{Zeynep Demir Vatansever$^2$}
\author{Erol Vatansever$^{1,2}$}
\author{Nikolaos G. Fytas$^1$}
\altaffiliation[]{Corresponding author: nikolaos.fytas@coventry.ac.uk}

\affiliation{$^1$Centre for Fluid and Complex Systems, Coventry
	University, Coventry, CV1 5FB, United Kingdom}

\affiliation{$^2$Department of Physics, Dokuz Eyl\"{u}l University, TR-35160, Izmir, Turkey}

\date{\today}

\begin{abstract}
We investigate by means of Monte Carlo simulations the dynamic phase transition of the two-dimensional kinetic Blume-Capel model under a periodically oscillating magnetic field in the presence of a quenched random crystal-field coupling. We analyze the universality principles of this dynamic transition for various values of the crystal-field coupling at the originally second-order regime of the corresponding equilibrium phase diagram of the model. A detailed finite-size scaling analysis indicates that the observed nonequilibrium phase transition belongs to the universality class of the equilibrium Ising ferromagnet with additional logarithmic corrections in the scaling behavior of the heat capacity. Our results are in agreement with earlier works on kinetic Ising models.
\end{abstract}

\maketitle

\section{Introduction}
\label{sec:introduction}

In the last decades our understanding of equilibrium critical
phenomena has developed to a point where well-established results
are available for a wide variety of systems. In particular, the
origin and/or the difference between equilibrium universality
classes is by now well understood. This observation also partially
holds for systems under the presence of quenched disorder.
However, far less is known for the physical mechanisms underlying
the nonequilibrium phase transitions of many-body interacting
systems that are far from equilibrium and clearly a solid classification
of nonequilibrium phase transitions into
universality classes is missing.

We know today that when a ferromagnetic system, below its Curie
temperature, is exposed to a time-dependent oscillating magnetic
field, it may exhibit a fascinating dynamical behavior~\cite{Tome}. 
In a typical ferromagnetic system being
subjected to an oscillating magnetic field, there occurs a
competition between the time scales of the half period of the applied field $t_{1/2}$
and the metastable lifetime, $\tau$, which is defined as the average time it takes the system to leave one of its two degenerate zero-field equilibrium states when a field of magnitude $h_{0}$ opposite to the initial magnetization is applied. In practice, $\tau$ is measured as the first-passage time to zero magnetization. When $t_{1/2} < \tau$, the time-dependent magnetization tends to oscillate around a nonzero value which corresponds to the dynamically ordered
phase. In this region, the time-dependent magnetization is not
capable of following the external field. However,
for larger values of the half period, the system
is given enough time to follow the external field and in this
case the time-dependent magnetization oscillates around its zero
value, indicating a dynamically disordered phase. When $t_{1/2} \approx \tau$, a dynamic phase transition takes place between the dynamically ordered and disordered phases.

Throughout the years, there have been several
theoretical~\cite{Lo, Zimmer, Acharyya1, Chakrabarti, Acharyya2,
	Acharyya3, Buendia1, Buendia2, Fujisaka, Jang1, Jang2, Shi, Punya,
	Riego, Keskin1, Keskin2, Robb1,Deviren, Yuksel1, Yuksel2,
	Vatansever1} and experimental
studies~\cite{He,Robb,Suen,Berger,Riego1} dealing with dynamic
phase transitions and hysteresis phenomena. The main conclusion  
is that both the amplitude and period of the time-dependent magnetic field play
a key role in dynamic critical phenomena. Furthermore, the characterization of
universality classes in spin models driven by a time-dependent
oscillating magnetic field has also attracted a lot of interest
lately~\cite{Sides1,Sides2,Korniss,Buendia3,Park,Park2,Tauscher,Buendia4,Vatansever_Fytas,Vatansever_Fytas2}. A short listing of the main results is given below:

(1) The critical exponents of the kinetic Ising model were found
	to be compatible to those of the equilibrium Ising model at both
	two (2D) and three dimensions
	(3D)~\cite{Sides1,Sides2,Korniss,Park}.
	
(2) Buend\'{i}a and Rikvold~\cite{Buendia3} estimated the critical exponents of the 2D Ising model and
	provided strong evidence that the characteristics of dynamic
	phase transition are universal with respect to the choice of
	stochastic dynamics. These authors used the so-called soft Glauber dynamics~\cite{Rikvold02},
	 for which both nucleation and interface propagation are slower and the interfaces smoother than for the standard hard Glauber and Metropolis dynamics.
	
(3) The role of surfaces at nonequilibrium phase transitions in
	Ising models has been elucidated by Park and Pleimling~\cite{Park2}: The
	nonequilibrium surface exponents were found to be different than
	their equilibrium counterparts.
	
(4) Experimental evidence by Riego \emph{et al.}~\cite{Riego1}
	and numerical results by Buend\'{i}a and Rikvold~\cite{Buendia4}
	verified that the equivalence of the dynamic phase transition to
	an equilibrium phase transition is limited to the area near the
	critical period and for zero bias.
	
(5) Numerical simulations by Vatansever and Fytas showed that
	the nonequilibrium phase transitions of the pure and random-bond spin-$1$ Blume-Capel
	model belong to the universality class of the equilibrium pure Ising model with logarithmic 
	corrections in the disordered case~\cite{Vatansever_Fytas,Vatansever_Fytas2}. Some general
	and very useful features of the dynamic phase transition of the pure
	Blume-Capel model can also be found in Refs.~\cite{Buendia1,Keskin1, Keskin2, Deviren, ShiWei, Acharyya4}.

The above results in 2D and 3D kinetic Ising and Blume-Capel
models establish a mapping between the universality principles of
the equilibrium and dynamic phase transitions of spin-$1/2$ and
spin-$1$ models. They also provide additional support in favor of an
earlier investigation of a Ginzburg-Landau model with a
periodically changing field~\cite{Fujisaka}, as well with the
symmetry-based arguments of Grinstein \emph{et al.} in
nonequilibrium critical phenomena~\cite{Grinstein}. 

As the vast majority of works in the field deal with pure systems, in the present paper we attempt to shed some additional light on the effect of quenched disorder on dynamic phase transitions~\cite{Vatansever_Fytas2}. We should note here some earlier mean-field and effective-field theory treatments of the problem where it was shown that the dynamic character of a typical system driven by a time-dependent magnetic field sensitively
depends on the amount of disorder, accounting for 
re-entrant phenomena and dynamic tricritical points~\cite{Gupinar12,Gupinar12b,Akinci12,Vatansever13,Vatansever13b,Vatansever15}.

In this paper we use as a test-case
platform for our numerical experiment the square-lattice Blume-Capel
model~\cite{Blume66} under a time-dependent magnetic
field, diffusing disorder in the crystal-field coupling 
[see below Eq.~(\ref{eq:bimodal})]. This type of randomness has also been used in the past for the equilibrium version of the model by Branco and Boechat~\cite{Branco97}, Sumedha and Mukherjee~\cite{Sumedha20}, Vatansever \emph{et al.}~\cite{Vatansever20} and is much closer to the experimental reality as it mimics the physics of random porous media in $^{3}$He --$^{4}$He mixtures~\cite{Buzano94}. In a nutshell, our extensive Monte-Carlo simulations for various values of the crystal-field coupling along the phase boundary indicate that the dynamic phase transition of the model belongs to the universality class of the corresponding equilibrium Ising model with logarithmic corrections in the heat-capacity scaling due to the presence of quenched disorder.

The rest of the paper is organized as follows: In
Sec.~\ref{sec:model_obs_method} we introduce the model, provide details of our simulation protocol and define the relevant thermodynamic observables. The numerical results are presented in Sec.~\ref{sec:results} and a summary of our conclusions is given in Sec.~\ref{sec:conclusions}.

\section{Model and Methods}
\label{sec:model_obs_method}

\subsection{Model}
\label{subsec:model}

The Hamiltonian of the Blume-Capel model under a time-dependent oscillating magnetic field reads as 
\begin{equation}\label{eq:Ham}
	\mathcal{H} = -J \sum_{\langle xy \rangle}\sigma_{x}\sigma_{y}+ \sum_{x}\Delta_{x}\sigma_{x}^2-h(t)\sum_{x}\sigma_{x},
\end{equation}
where the spin variable $\sigma_{x}$ takes on the
values $\{-1,0,+1\}$, $\langle xy \rangle$ indicates summation over
nearest neighbors on the square lattice and the coupling $J > 0$ denotes the
ferromagnetic exchange interactions. $\Delta_{x}$ represents the crystal-field strength and controls the density of vacancies ($\sigma_{x} = 0$). As mentioned above we choose a site-dependent bimodal crystal-field probability distribution of the form
\begin{equation} \label{eq:bimodal}
	\mathcal{P}(\Delta_{x})=p\delta(\Delta_{x}+\Delta)+(1-p)\delta(\Delta_{x}-\Delta),
\end{equation}
where $p\in (0,1)$ is the control parameter of the disorder distribution with $\mu = \Delta (1-2p)$ and $s = 2\Delta \sqrt{p(1-p)}$ the mean value and standard deviation of the distribution~(\ref{eq:bimodal}), respectively. Finally, the term $h(t)$ corresponds to a spatially uniform periodically oscillating magnetic field, so that all
lattice sites are exposed to a square-wave magnetic field with
amplitude $h_{0}$ and half period $t_{1/2}$~\cite{Korniss,Buendia3,Park}.

Some useful explanatory comments for the equilibrium ($h(t) = 0$) version of the model are in order: 

(1) For $\Delta = \infty $ the model is equivalent to the random site spin-$1/2$ Ising model, where sites are present or absent with probability $p$ or $1-p$, respectively~\cite{Branco97}. 

\begin{figure}
	\includegraphics[width=11 cm]{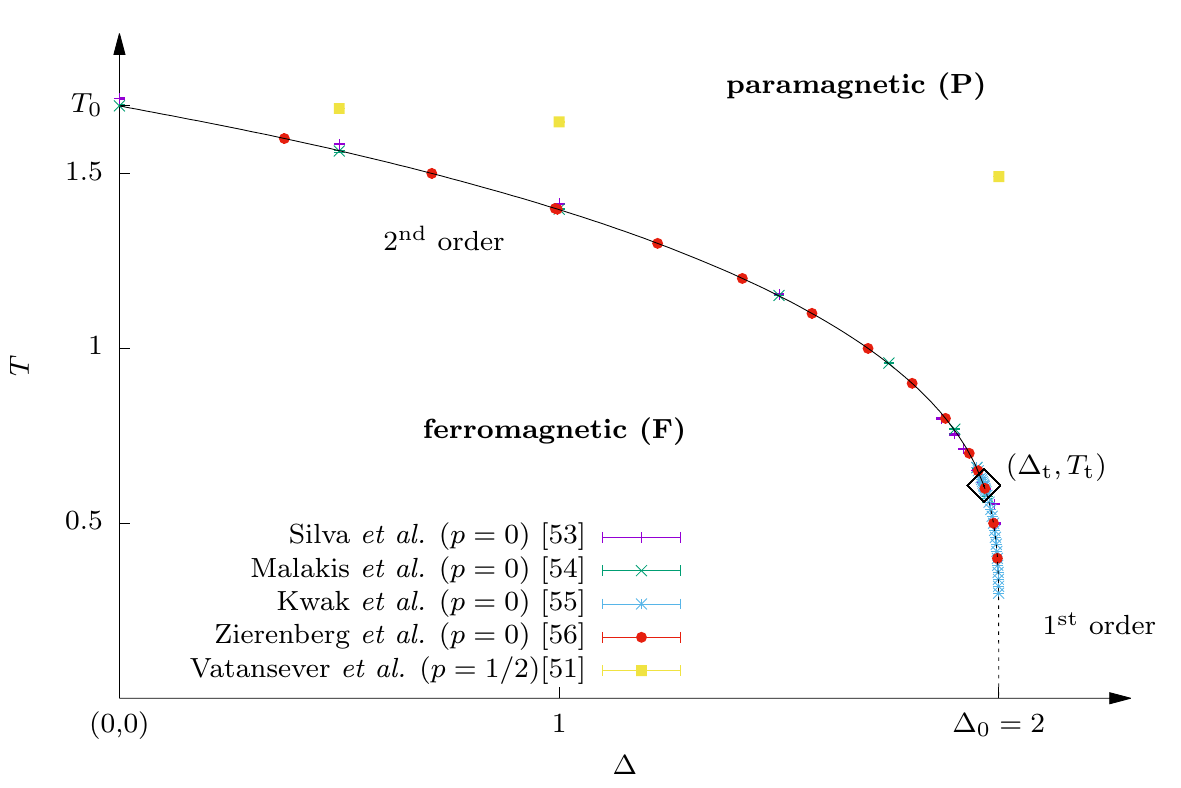}
	\caption{Phase diagram of the pure ($p=0$) and random ($p=1/2$) square-lattice
		Blume-Capel model in the $\Delta$ -- $T$ plane showing the ferromagnetic
		(\textbf{F}) and paramagnetic (\textbf{P}) phases that are
		separated by a continuous transition at small $\Delta$ (solid
		line) and a first-order at large $\Delta$ (dotted
		line). The line segments meet at a tricritical point ($\Delta_{\rm t}$, $T_{\rm t}$) marked
		by a black rhombus. Numerical data shown are selected estimates from previous studies, as indicated also in the panel.
		\label{fig:diagram}}
\end{figure}

(2) For $p = 0$ the pure Blume-Capel model is recovered~\cite{Silva06,Malakis10,Kwak15,Zierenberg17}. The phase diagram of the pure ($p=0$) and random ($p=1/2$) model in the $\Delta$ -- $T$ plane is shown
in Fig.~\ref{fig:diagram} including a variety of critical and transition points from the current literature. For small $\Delta$ there is a line of continuous transitions (in the Ising universality class)
between the ferromagnetic and paramagnetic phases that crosses the $\Delta = 0$ axis
at $T_{0}\approx 1.693$~\cite{Malakis10}. For large $\Delta$ the
transition becomes discontinuous and it meets the $T=0$ line at
$\Delta_0 = zJ/2$~\cite{Blume66}, where $z = 4$ is the coordination number (as usual we
set $J = k_{\rm B} = 1$ to fix the temperature scale). The two line
segments meet at a tricritical point
$(\Delta_{\rm t}\approx1.966,T_{\rm t}\approx0.608)$~\cite{Kwak15}. 

(3) With the inclusion of disorder ($p > 0$) the critical temperature of the system rises -- see the yellow filled squares in Fig.~\ref{fig:diagram}. For further explanations and simple arguments explaining this behavior we refer the reader to Ref.~\cite{Vatansever20}.

\subsection{Numerical approach}
\label{subsec:method}

We performed Monte Carlo simulations with
periodic boundary conditions using the single-site update
Metropolis algorithm~\cite{Metropolis,Binder,Newman}. This
approach, together with the stochastic Glauber
dynamics~\cite{Glauber63}, consists the standard recipe in
kinetic Monte Carlo simulations~\cite{Buendia3}. Let us briefly outline below the steps of our algorithm:

(1) A lattice site is selected randomly among the $N = L \times L$ options.

(2) The spin variable located at the selected site is flipped, keeping the other spins in the system fixed. 

(3) The energy change originating from this spin flip operation is calculated using the Hamiltonian of Eq.~(\ref{eq:Ham}) via $\Delta \mathcal{H}=\mathcal{H}_{\rm a}-\mathcal{H}_{\rm o}$, where $\mathcal{H}_{\rm a}$ denotes the system’s energy after the trial switch of the selected spin and $\mathcal{H}_{\rm o}$ corresponds to the total energy of the old spin configuration. The probability to accept the proposed spin update is given by
\begin{equation}
	\label{eq:prob_accept}
	\Pi\left(\sigma_{x}\rightarrow \sigma_{x}' \right)=
	\begin{cases}
		\exp(-\Delta \mathcal{H}/k_{\rm B}T),        & \text{if } \mathcal{H}_{a} \geq \mathcal{H}_{o} \\
		1,        & \text{if } \mathcal{H}_{\rm a} < \mathcal{H}_{\rm o}.
	\end{cases}
\end{equation}

(4) If the energy is lowered, the spin flip is always accepted. 

(5) If the energy is increased, a random number $R$ is generated, such that $0 \leq  R < 1$: If $R$ is less than or equal to the calculated Metropolis transition probability the selected spin is flipped. Otherwise, the old spin configuration remains unchanged. Note that all transitions among the three spin states $\{-1,0,+1\}$ are allowed in our numerical protocol.

Using the above scheme we simulated the model of Eqs.~(\ref{eq:Ham}) and (\ref{eq:bimodal}) at $\Delta=0.5$, $1$, and $\Delta = 2$, fixing the control parameter $p$ to the value $1/2$, guided by the analysis of Ref.~\cite{Vatansever20}. System sizes varied within the range $L  = 32 - 512$ and for each linear size an average over $500$ independent realizations of the disorder was performed. The first $10^{3}$ periods of the external field were discarded during the thermalization process and numerical data were collected and analyzed during the following $11\times 10^{3}$ periods of the field. Note that the time unit in our simulations is one Monte Carlo step per site (MCSS) and that error bars were estimated using the jackknife method~\cite{Newman}. Appropriate choices of magnetic-field strength, $h_{0} = 0.3$, and temperature, $T (\Delta) =0.8\times T_{\rm c}(\Delta)$, ensured that the metastable decay of the system following field reversal occurs through nucleation and growth of many droplets of the stable phase, \emph{i.e.}, the multidroplet regime. This point was already emphasized by Sides \emph{et al.} in 1998~\cite{Sides1} -- see also Ref.~\cite{Park}. Here, $T_{\rm c}(\Delta = 0.5) =  1.6854$, $T_{\rm c}(\Delta = 1) =  1.6473$, and $T_{\rm c}(\Delta = 2) =  1.4907$ are the equilibrium critical
temperatures of the random $p=1/2$ Blume-Capel model defined in Eqs.~(\ref{eq:Ham}) and (\ref{eq:bimodal})~\cite{Vatansever20}.

Finally a comment on the fitting process discussed below in Sec.~\ref{sec:results}: We employed the
standard $\chi^{2}$ goodness of fit test~\cite{Press}. Specifically, the $Q$-value of our $\chi^{2}$-test is the probability of finding a $\chi^{2}$ value which is even larger
than the one actually found from our data. We consider a fit as being fair only if $10\% \leq Q \leq 90\%$.

\subsection{Observables}
\label{subsec:observables}

In order to determine the universality aspects of the kinetic
random Blume-Capel model, we consider the
half-period dependencies of various thermodynamic observables. The
main quantity of interest is the period-averaged magnetization
\begin{equation}\label{eq:order_parameter}
	Q_{L}=\frac{1}{2t_{1/2}}\oint M(t)dt,
\end{equation}
where the integration is performed over one cycle of the
oscillating field. Given that for finite systems in the
dynamically ordered phase the probability density of $Q_{L}$ becomes
bimodal, one has to measure the average norm of $Q_{L}$ in order to
capture symmetry breaking so that $\langle |Q| \rangle_{L}$ defines
the dynamic order parameter of the system. In
Eq.~(\ref{eq:order_parameter}), $M(t)$ is the time-dependent magnetization per
site
\begin{equation}\label{eq:magnetization}
	M(t)=\frac{1}{N}\sum_{x = 1}^{N}\sigma_{x}(t).
\end{equation}

To characterize and quantify the transition using finite-size
scaling arguments we must also define quantities analogous to the
susceptibility in equilibrium systems. The scaled variance of the
dynamic order parameter
\begin{equation}\label{eq:mag_susceptibility}
	\chi_{L}^{Q} = N\left[\langle Q^2\rangle_{L} -\langle |Q|
	\rangle^2_{L} \right]
\end{equation}
has been suggested as a proxy for the nonequilibrium
susceptibility, also theoretically justified via
fluctuation-dissipation relations~\cite{Robb1}. 

Similarly, one may
also measure the scaled variance of the period-averaged energy
\begin{equation} \label{eq:spec_heat}
	\chi_{L}^{E} = N\left[\langle E^2\rangle_{L} -\langle E
	\rangle^2_{L} \right],
\end{equation}
so that $\chi_{L}^{E}$ can be considered as the respective heat capacity. 
Here $E$ denotes the
cycle-averaged energy corresponding to the cooperative part of the
Hamiltonian~(\ref{eq:Ham}). With the help of $Q$ we may also define the fourth-order Binder cumulant~\cite{Sides1,Sides2}
\begin{equation}
	\label{eq:binder} U_{L}^{Q}=1-\frac{\langle |Q|^4\rangle_L}{3\langle |Q|^2
		\rangle_L^2},
\end{equation}
a very useful observable for the characterization of universality
classes~\cite{Binder81}.

\section{Results}
\label{sec:results}

As a starting point let us describe shortly the mechanism
underlying dynamic ordering in kinetic ferromagnets as depicted in
Figs.~\ref{fig:timeseries} - \ref{fig:configurations} below. In all these plots
results for a single realization of the disorder are shown of a
system size $L = 192$ and for $\Delta = 1$. Similar results were obtained also for the other $\Delta$ values but are omitted for brevity.

Figure~\ref{fig:timeseries} presents the time evolution of
the magnetization and Fig.~\ref{fig:period} the period
dependencies of the dynamic order parameter $Q$ of the kinetic
random Blume-Capel model. For rapidly varying fields, Fig.~\ref{fig:timeseries}(a), the
magnetization does not have enough time to switch during a single
half period and remains nearly constant for many successive field
cycles, as also illustrated by the black line in
Fig.~\ref{fig:period}. On the other hand, for slowly varying
fields, Fig.~\ref{fig:timeseries}(c), the magnetization
follows the field, switching every half period, so that $Q \approx 0$, as also shown by the blue line in
Fig.~\ref{fig:period}. Thus, whereas in the
dynamically disordered phase the ferromagnet is able to reverse
its magnetization before the field changes again, in the
dynamically ordered phase this is not possible and therefore the
time-dependent magnetization oscillates around a finite value. The
competition between magnetic field and the metastable state is
captured by the half period $t_{1/2}$ (or by the
normalized parameter $\Theta = t_{1/2} /\tau$~\cite{Park}). Obviously, $t_{1/2}$ plays
the role of temperature in the equilibrium system. Now, the
transition between the two regimes is characterized by strong
fluctuations in $Q$, see Fig.~\ref{fig:timeseries}(b) and the evolution of the red line
in Fig.~\ref{fig:period}. This behavior is indicative of a
dynamic phase transition and occurs for values of the half period
close to the critical one $t_{1/2}^{\rm c}$ (otherwise when $\Theta \approx 1$). 
Of course, since the value $t_{1/2} = 66$ MCSS used for this illustration is
slightly above $t^{\rm c}_{1/2} = 65.96(6)$, see also
Fig.~\ref{fig:pseudocritical} below, the observed behavior includes as well some
nonvanishing finite-size effects. 

Some additional spatial aspects of the transition scenarios
described above via the configurations of a local order
parameter $\{Q_{x}\}$ are shown in Fig.~\ref{fig:configurations}. 
Below $t_{1/2}^{\rm c}$, see panel (a), the majority of spins
spend most of their time in the $+1$ state, \emph{i.e.}, in the
metastable phase during the first half period, and in the stable
equilibrium phase during the second half period, except for
fluctuations. Thus, most of the $Q_{x}\approx +1$ and
the system lies in the dynamically ordered phase. On the other
hand, when the period of external field is selected to be
bigger than the relaxation time of the system, above $t_{1/2}^{\rm c}$, see panel (c), the system follows the field in every half period with some phase lag, and $Q_{x}\approx 0$ at all sites
$x$. The system in this case is in the dynamically disordered phase. Near
$t_{1/2}^{\rm c}$ and the expected dynamic phase transition, there
are large clusters of both $Q_{x}\approx +1$ and $-1$ values
within a sea of $Q_{x}\approx 0$, as shown in
panel (b).

At this point we would like to scrutinize the effects of the zero spin state $\sigma_{x} = 0$ and (random) crystal-field coupling $\Delta$, in comparison to the well established picture of the standard Ising ferromagnet. Although there is no doubt that the local order parameter of most interest is $\{Q_{x}\}$, yet, it can not distinguish between random distributions of $\sigma_{x} = \pm 1$ and regions of $\sigma_{x} = 0$. To bring out this distinction, we present in Fig.~\ref{fig:configurations_new}  configurations of the dynamic quadrupole moment $\{O_{x}\}$ over a full cycle of the external
field, where $O = \frac{1}{2t_{1/2}}\oint \rho(t)dt$ and
$\rho(t)= 1 - \frac{1}{N}\sum_{x=1}^{N}\sigma_{x}^2$ denotes the order parameter conjugate to the crystal-field coupling $\Delta$. Moreover, in analogy to Figs.~\ref{fig:timeseries} and \ref{fig:period}, the additional Figs.~\ref{fig:timeseries_new} and \ref{fig:period_new} present the time evolution of $\rho(t)$ and the period dependencies of the quadrupole moment of the kinetic
random $p=1/2$ Blume-Capel model. In all Figs.~\ref{fig:configurations_new} -- \ref{fig:period_new} simulation parameters are exactly the same to those used in Figs.~\ref{fig:timeseries} -- \ref{fig:configurations} above. 

Of course, the dynamic quadrupole moment is always $0$ for the kinetic spin-1/2 Ising model, because $\sigma_{x} = \pm 1$ in this case. For the spin-$1$ Blume-Capel model
the density of vacancies is controlled by the crystal-field coupling $\Delta$ and, thus, the value of
$O$ changes depending on $\Delta$. When
the  value of $\Delta$ increases, starting from the Ising limit
$(\Delta \rightarrow -\infty)$, the number of vacancies increases
as well in the system, so that $O$ tends to increase from its minimum value. For the case of the kinetic random $p=1/2$ Blume-Capel model at $\Delta = 1$, as depicted in Figs.~\ref{fig:configurations} and \ref{fig:configurations_new}, one may conclude that the effect of vacancies is not significant. Moreover, for this particular case of $\Delta = 1$ we have performed a quantitative comparison among the pure ($p=0$) and random ($p=1/2$) model and did not observe any significant differences in the configurations of the local dynamic order parameter $Q$ and quadrupole moment $O$ that are worth to be noted. On the other hand, we expect to see more prominent effects in the small-$p$ and high-$\Delta$ limits that correspond to the ex-first-order transition regime of the equilibrium model's phase diagram~\cite{Vatansever20}. In fact, the set of parameters $p = 0.02$ and $\Delta = 2$ may be a promising choice and we present in Figs.~\ref{fig:configurations_low_p} and \ref{fig:configurations_low_p_new} three sets of configurations for both the local dynamic order parameter and quadrupole moment, below, around, and above the dynamic phase transition. These snapshots fully corroborate our claim that in this regime the underlying phenomena are indeed controlled by the vacancies, as expected.

To further explore the nature of dynamic phase transitions
encountered in the above disordered kinetic model we performed a
finite-size scaling analysis based on the observables outlined in
Sec.~\ref{subsec:observables}. Previous studies in the field
indicated that although finite-size scaling is a
tool that has been designed for the study of equilibrium phase
transitions, it can be successfully applied as well to far from
equilibrium systems~\cite{Sides1,Sides2,Korniss,Buendia3,Park}.

As an illustrative example we present in Fig.~\ref{fig:curves_magnetic} the finite-size behavior of the dynamic order parameter (main panel) and the emerging
susceptibility (inset) -- see also Eq.~(\ref{eq:mag_susceptibility}) -- for the case $\Delta = 1$ and for two characteristic system sizes. The dynamic
order parameter starts off from a finite value and approaches zero as the
half period increases, showing a sharp change for a range of $t_{1/2}$ values 
that correspond to the respective peak in  
the dynamic susceptibility. These maxima locations of
$\chi_{L}^{Q}$, denoted hereafter as $(\chi^{Q}_{L})^{\ast}$, may be used to define suitable pseudocritical half periods $t_{1/2}^{\ast}$. In full analogy 
we may also denote the heat-capacity maxima as
$(\chi_{L}^{E})^{\ast}$. 

We start the presentation of our finite-size scaling analysis with
a characteristic determination of the critical half period $t_{1/2}^{\rm c}$ and the exponent $\nu$ for the system with $\Delta = 1$. A similar
analysis was performed for the other values of $\Delta$ as well and a summary of our findings
is given in Tab.~\ref{table}. The main panel of Fig.~\ref{fig:pseudocritical} illustrates the
shift behavior of the peak locations $t_{1/2}^{\ast}$ of the dynamic susceptibility and heat capacity as a function of $1/L$. The solid lines
are a joint fit of the form~\cite{Fisher,Privman,Binder92}
\begin{equation} \label{eq:shift}
	t_{1/2}^{\ast} = t_{1/2}^{\rm c} + bL^{-1/\nu}.
\end{equation}
The obtained values for the critical parameters are $t_{1/2}^{\rm c} = 65.96(6)$ and $\nu =
1.03(3)$. Clearly, the value of $\nu$ is in very good
agreement with the value of $\nu = 1$ in the 2D equilibrium Ising
universality class~\cite{Ferdinand69}.

Additional evidence of universality may be obtained from the
fourth-order Binder cumulant $U_{L}^{Q}$ defined in Eq.~(\ref{eq:binder}) 
for the case of the dynamic order parameter. In the inset of Fig.~\ref{fig:pseudocritical}
we present our numerical data of $U_{L}^{Q}$ for $\Delta = 1$ and a wide range of sizes studied.
The vertical dashed line marks the critical
half-period value of the system $t_{1/2}^{\rm c}$ as estimated from the analysis of Eq.~(\ref{eq:shift}) and the horizontal
dashed line the universal value $U^{\ast} = 0.610\;6924(16)$
of the 2D equilibrium Ising model~\cite{Salas}. Certainly, the crossing point 
is expected to depend on the lattice size $L$ (as it is also shown in the figure) 
and the term universal is valid for given lattice shapes, boundary conditions, and isotropic
interactions~\cite{Selke_2,Selke_3}. However, the data shown in the inset of Fig.~\ref{fig:pseudocritical} support,
at least qualitatively, another instance of equilibrium Ising universality, since the 
crossing point is consistent to the value $0.610\;6924$. 
We should note here that Hasenbusch \emph{et al.}
presented very strong evidence that the critical Binder cumulant of the equilibrium 
2D randomly site-diluted Ising model maintains its pure-system value~\cite{Hasenbusch08}.  
In this respect, a dedicated study along the lines of Ref.~\cite{Hasenbusch08} for an 
accurate estimation of $U^{\ast}$ in kinetic random Ising and Blume-Capel models
would be welcome.

In this final part we investigate the finite-size scaling behavior of the dynamic susceptibility and heat-capacity maxima. In particular we present in Fig.~\ref{fig:chi} the size evolution
of the dynamic susceptibility peaks in a log-log scale for all three values of $\Delta$ considered. The solid lines are a fit of the form~\cite{Ferrenberg91}
\begin{equation}
	\label{eq:chi_scaling}
	(\chi^{Q}_{L})^{\ast} \sim L^{\gamma/\nu},
\end{equation}
providing estimates for the magnetic exponent ratio $\gamma/\nu$ in excellent agreement with the Ising universality class value of $7/4$ -- see also Tab.~\ref{table} below. At this stage, it would be ideal to also observe the possible double logarithmic scaling behavior of the heat-capacity maxima $(\chi_{L}^{E})^{\ast}$, as predicted by Ref.~\cite{Dotsenko81} for the disordered Ising
ferromagnet. Indeed, as it is shown in
 Fig.~\ref{fig:sp_heat} the data for $L\geq 64$ are fairly good described by a fit of the form
\begin{equation}
	\label{eq:spe_heat_scaling} 
	(\chi^{E}_{L})^{\ast} \sim \ln{[\ln{(L)}]}.
\end{equation}
 
\section{Conclusions}
\label{sec:conclusions}

We investigated, using extensive Monte Carlo simulations, the effect of quenched disorder in the crystal-field coupling on the dynamic phase transition of the square-lattice Blume-Capel model under a periodically oscillating magnetic field. At a first qualitative level, the role of vacancies and the crystal-field coupling has been scrutinized by examining the configurations of the dynamic order parameter and quadrupole moment of the system for a wide range of simulation parameters. At a second stage, the application of finite-size scaling techniques allowed us to probe with good accuracy the values of critical exponents describing this dynamic phase transition, all of which were found to be compatible with those of the equilibrium Ising ferromagnet. An additional study of the scaling behavior of the heat capacity revealed a double logarithmic divergence, as
expected for the disordered Ising ferromagnet. To conclude, although universality is a cornerstone in the theory of critical phenomena it stands on a less solid foundation for the case of nonequilibrium systems under the presence of quenched disorder. We hope that our contribution will stimulate further research in this direction.

\begin{acknowledgments}
The authors would like to thank the two anonymous referees for their instructive comments. The numerical calculations reported in this paper were performed at T\"{U}B\.{I}TAK ULAKB\.{I}M (Turkish agency), High Performance and Grid Computing Center (TRUBA Resources).  We also acknowledge the provision of computing time on the parallel computer cluster \emph{Zeus} of Coventry University.
\end{acknowledgments}

\newpage

\begin{figure}[t]
	\centering
	\includegraphics[width=8 cm]{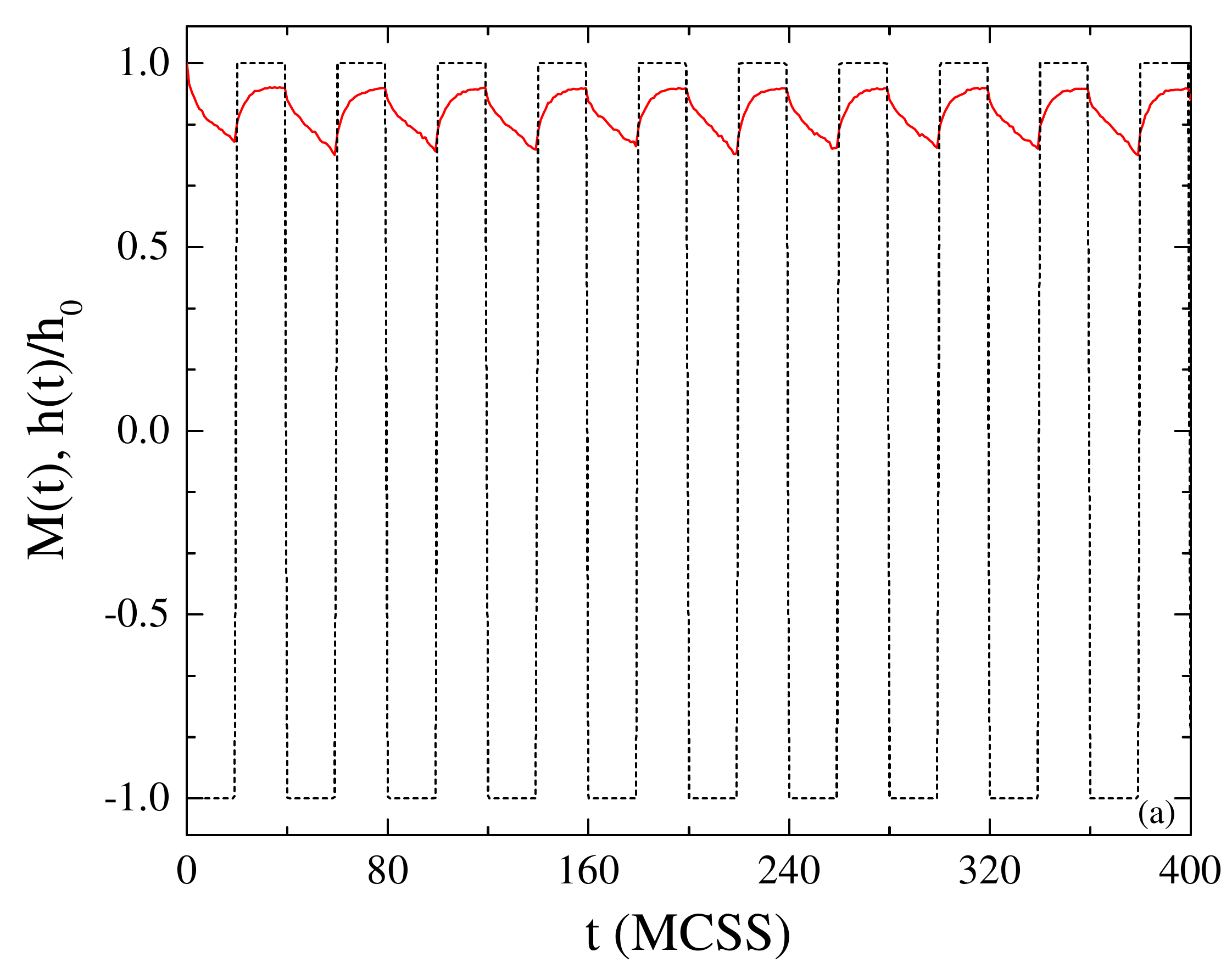}\\
	\includegraphics[width=8 cm]{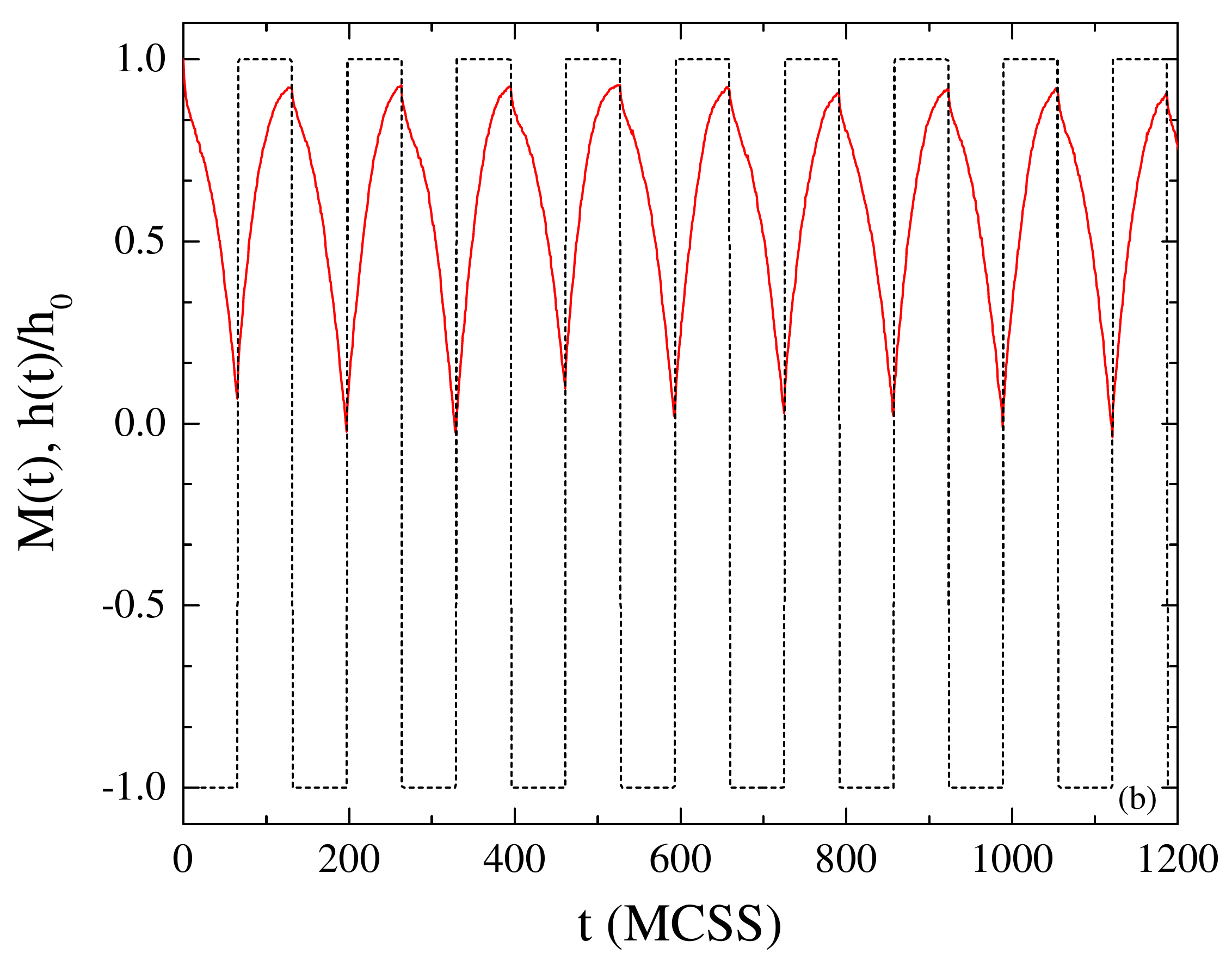}\\
	\includegraphics[width=8 cm]{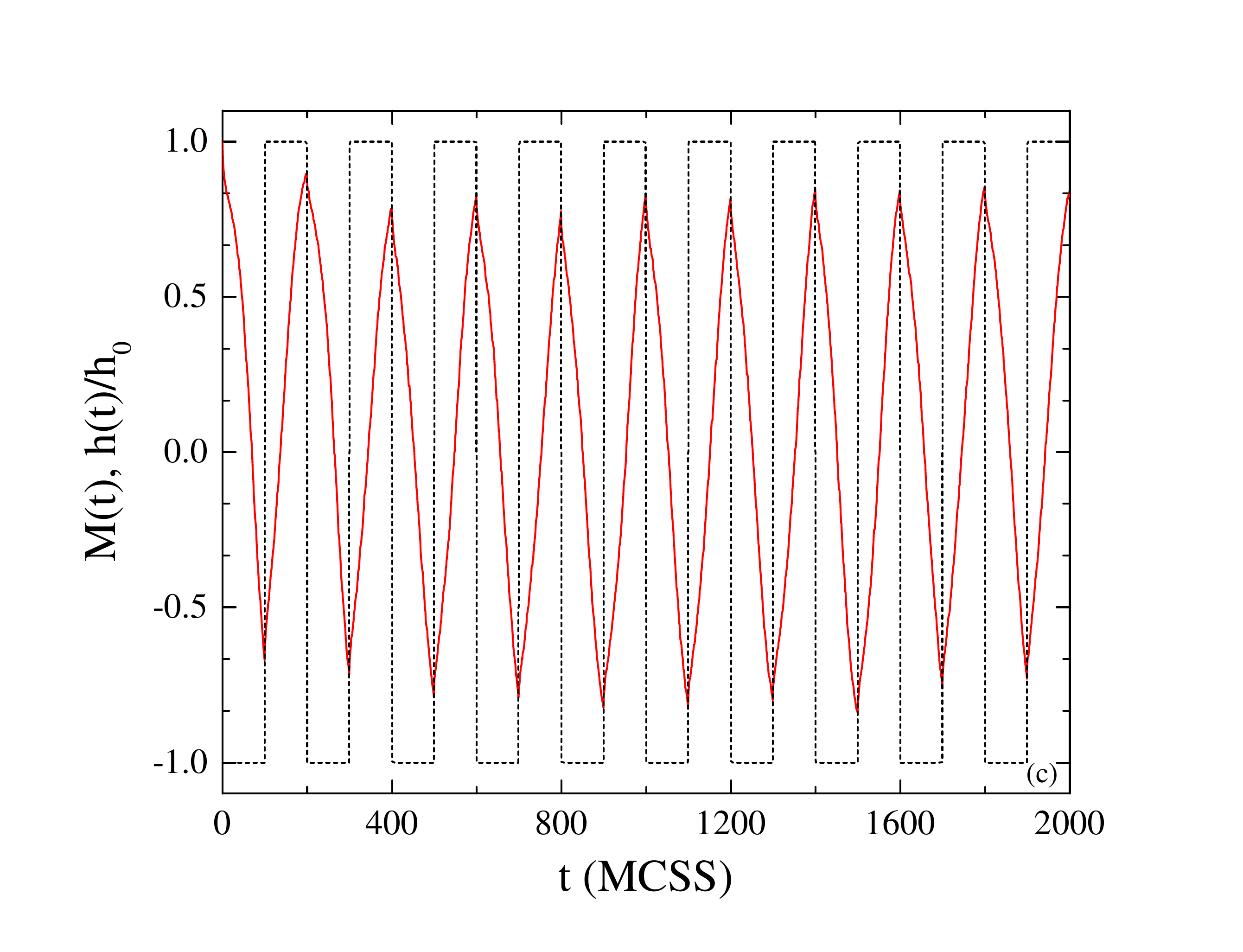}
	\caption{\label{fig:timeseries} Time series of the
		magnetization (red solid curves) of the kinetic random $p=1/2$ Blume-Capel model under the presence of a square-wave magnetic field (black
		dashed lines) for $L = 192$ at $\Delta = 1$, for three values of the half period of
		the external field: (a) $t_{1/2} = 20$ MCSS, corresponding to a
		dynamically ordered phase, (b) $t_{1/2} = 66$ MCSS, close to the
		dynamic phase transition, and (c) $t_{1/2} = 100$ MCSS,
		corresponding to a dynamically disordered phase. Note that for the
		sake of clarity the ratio $h(t)/h_{0}$ is displayed.}
\end{figure}

\begin{figure}
	\includegraphics[width=11 cm]{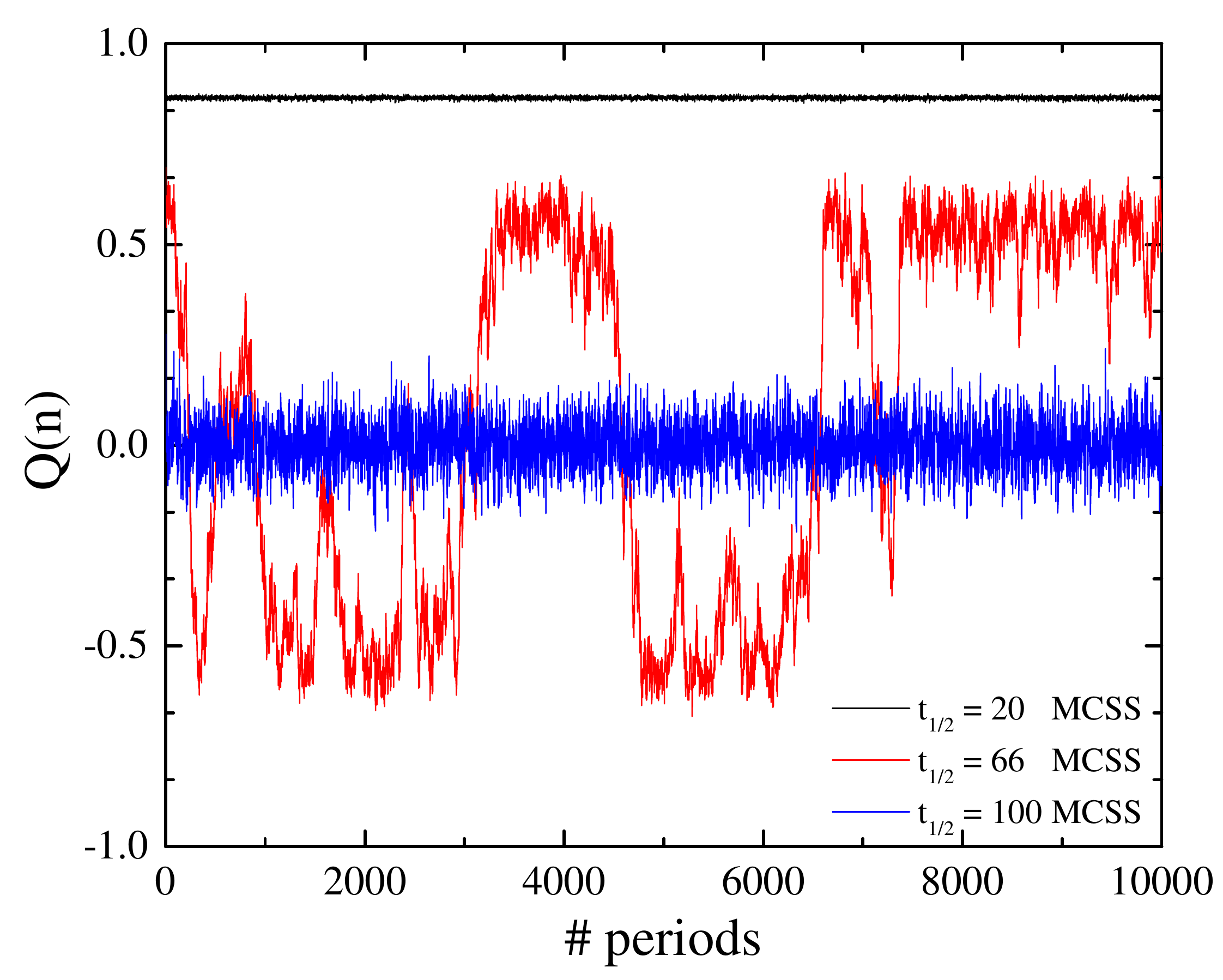}
	\caption{Period dependencies of the
		dynamic order parameter of the kinetic random $p=1/2$ Blume-Capel model for
		$L = 192$ at $\Delta = 1$. Results are shown for the three characteristic cases of
		the half period of the external field, following
		Fig.~\ref{fig:timeseries}.
		\label{fig:period}}
\end{figure}

\begin{figure}[t]
	\centering
	\includegraphics[width=8 cm]{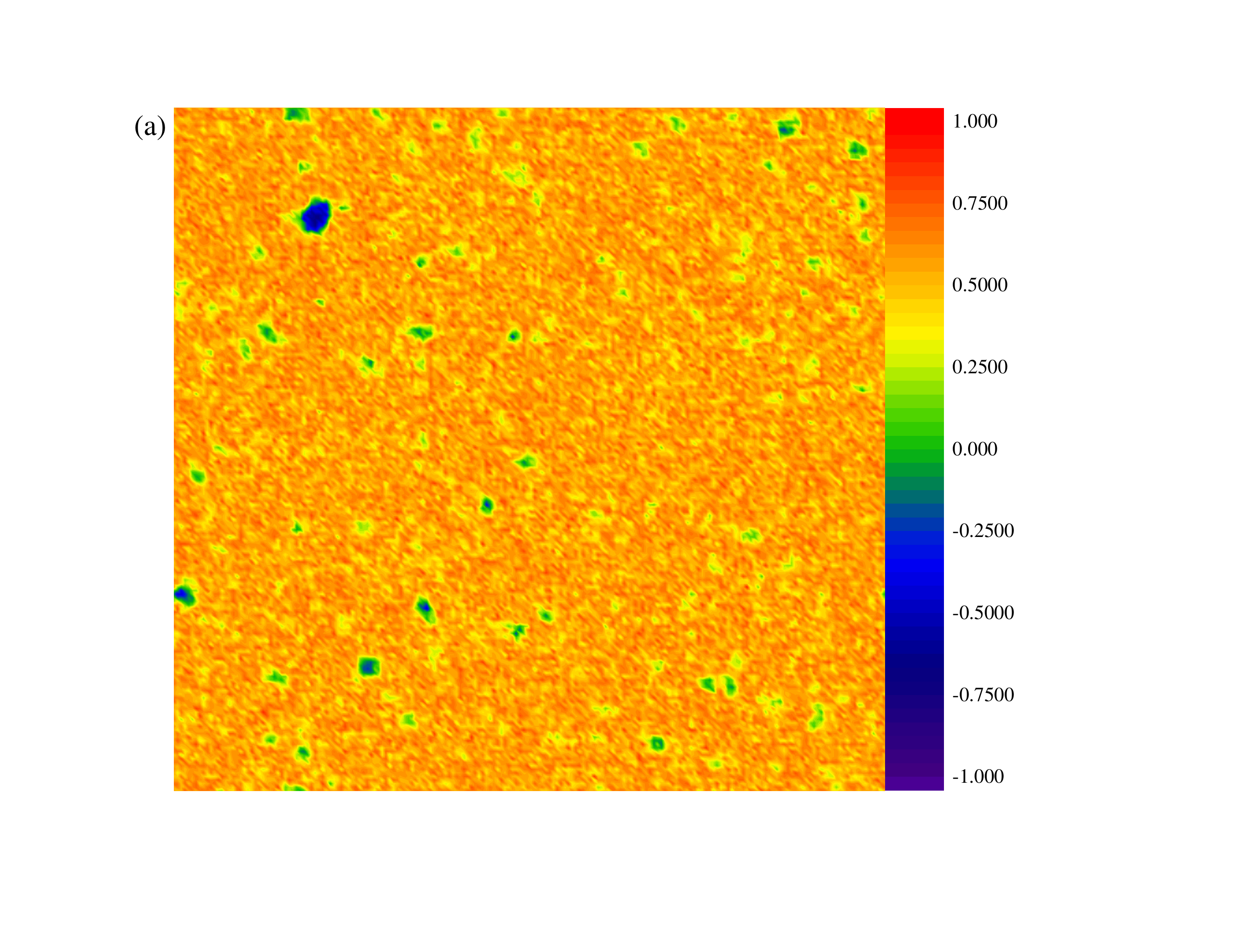}\\
	\includegraphics[width=8 cm]{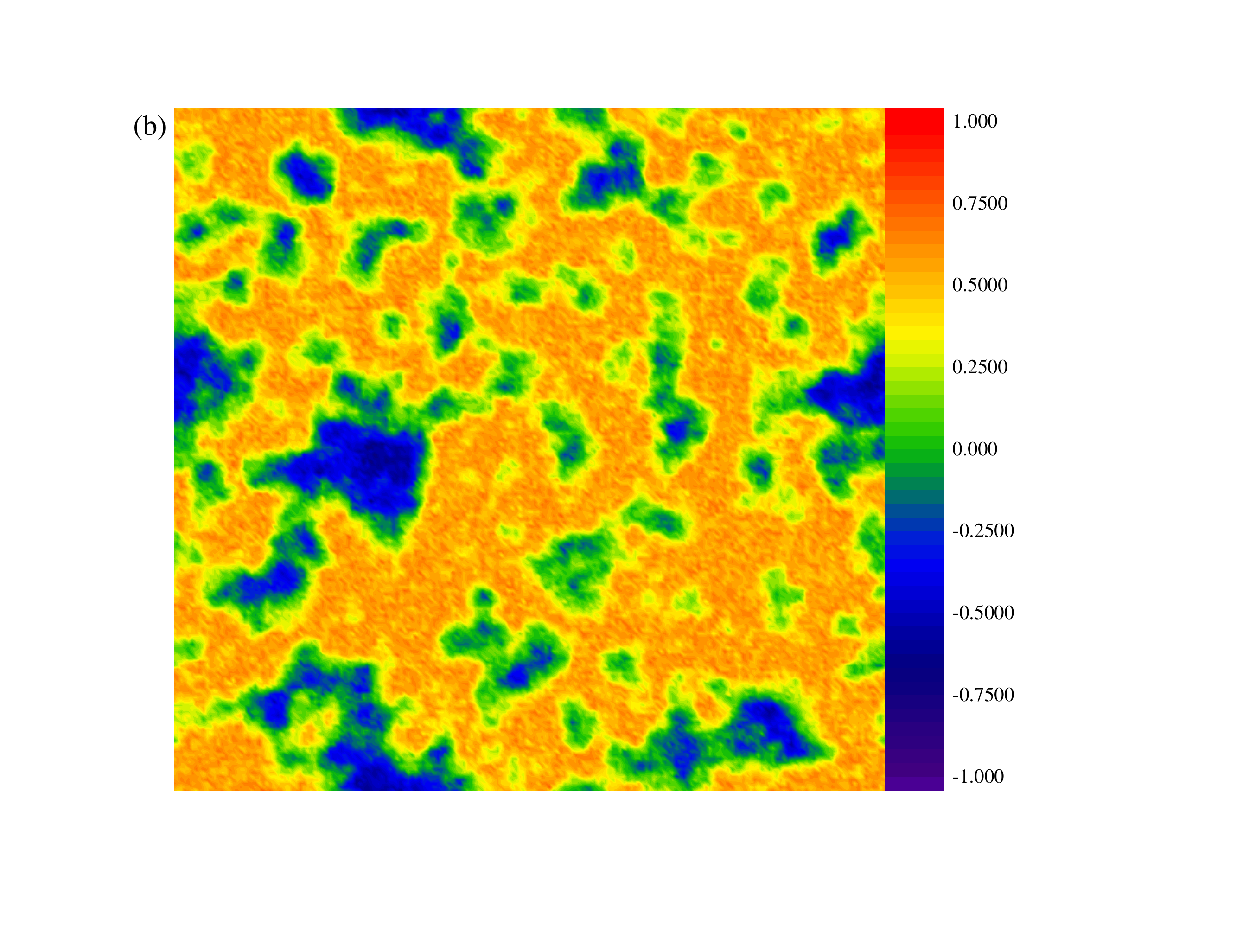}\\
	\includegraphics[width=8 cm]{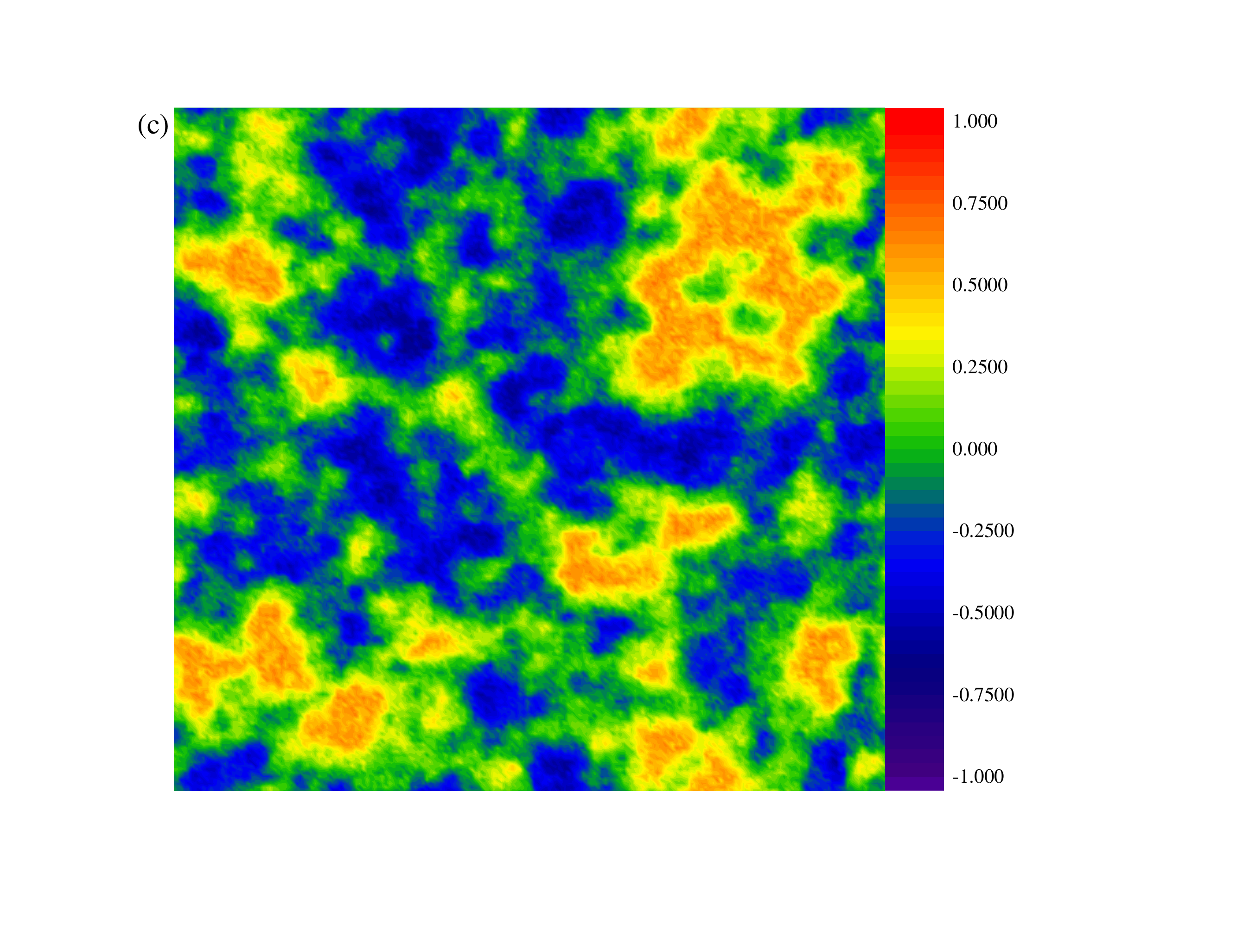}
	\caption{\label{fig:configurations} Configurations of the local
		dynamic order parameter $\{Q_{x}\}$ of the kinetic random $p=1/2$ Blume-Capel
		model for $L = 192$ at $\Delta = 1$. The ``snapshots'' of $\{Q_{x}\}$ for
		each regime are the set of local period-averaged spins during some
		representative period. Three panels are shown: (a) $t_{1/2} = 20$
		MCSS $ < t_{1/2}^{\rm c}$ -- dynamically ordered phase, (b)
		$t_{1/2} = 66$ MCSS $\approx t_{1/2}^{\rm c}$ -- near the dynamic
		phase transition, and (c) $t_{1/2} = 100$ MCSS $ > t_{1/2}^{\rm
			c}$ -- dynamically disordered phase.}
\end{figure}


\begin{figure}[t]
	\centering
	\includegraphics[width=8 cm]{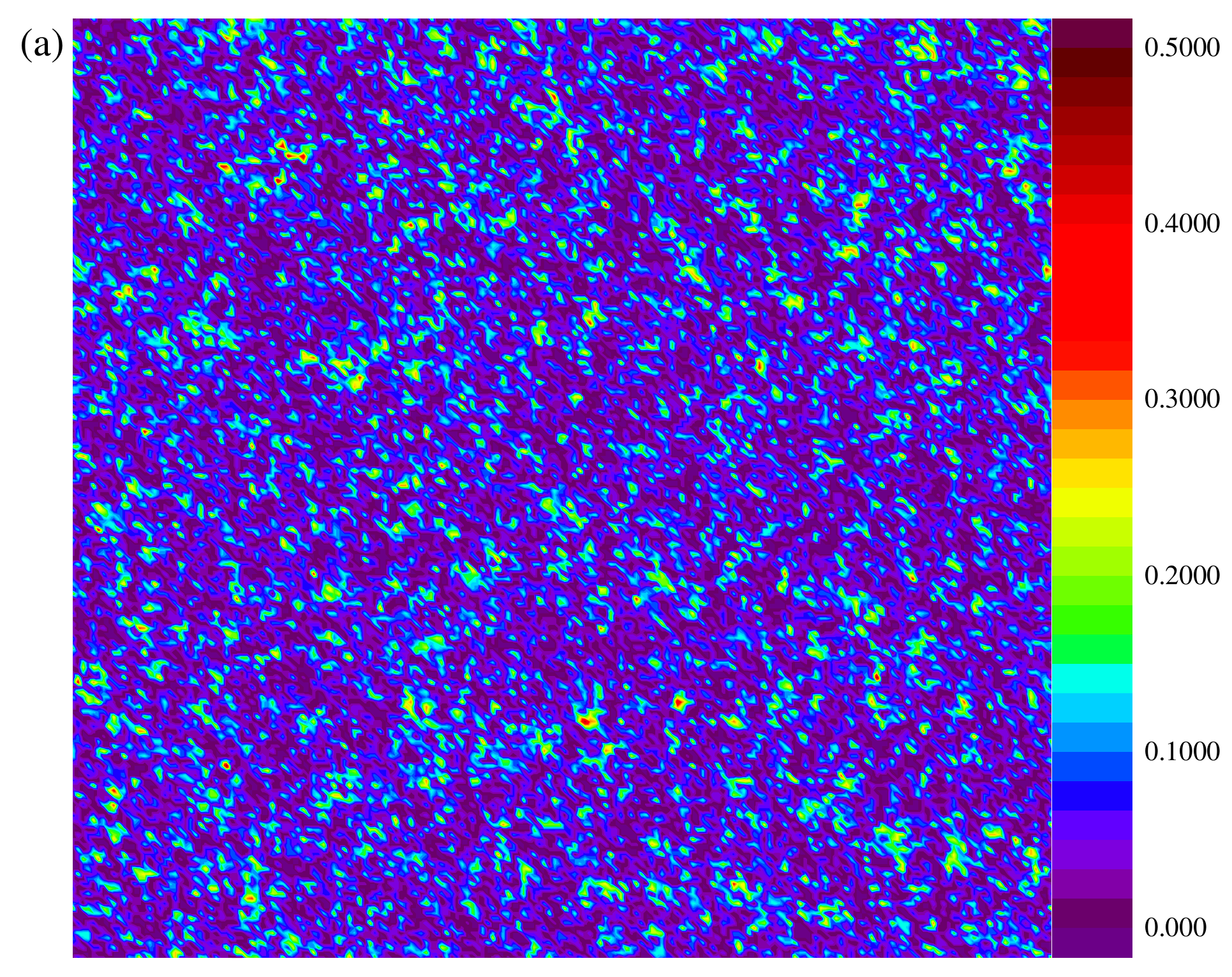}\\
	\includegraphics[width=8 cm]{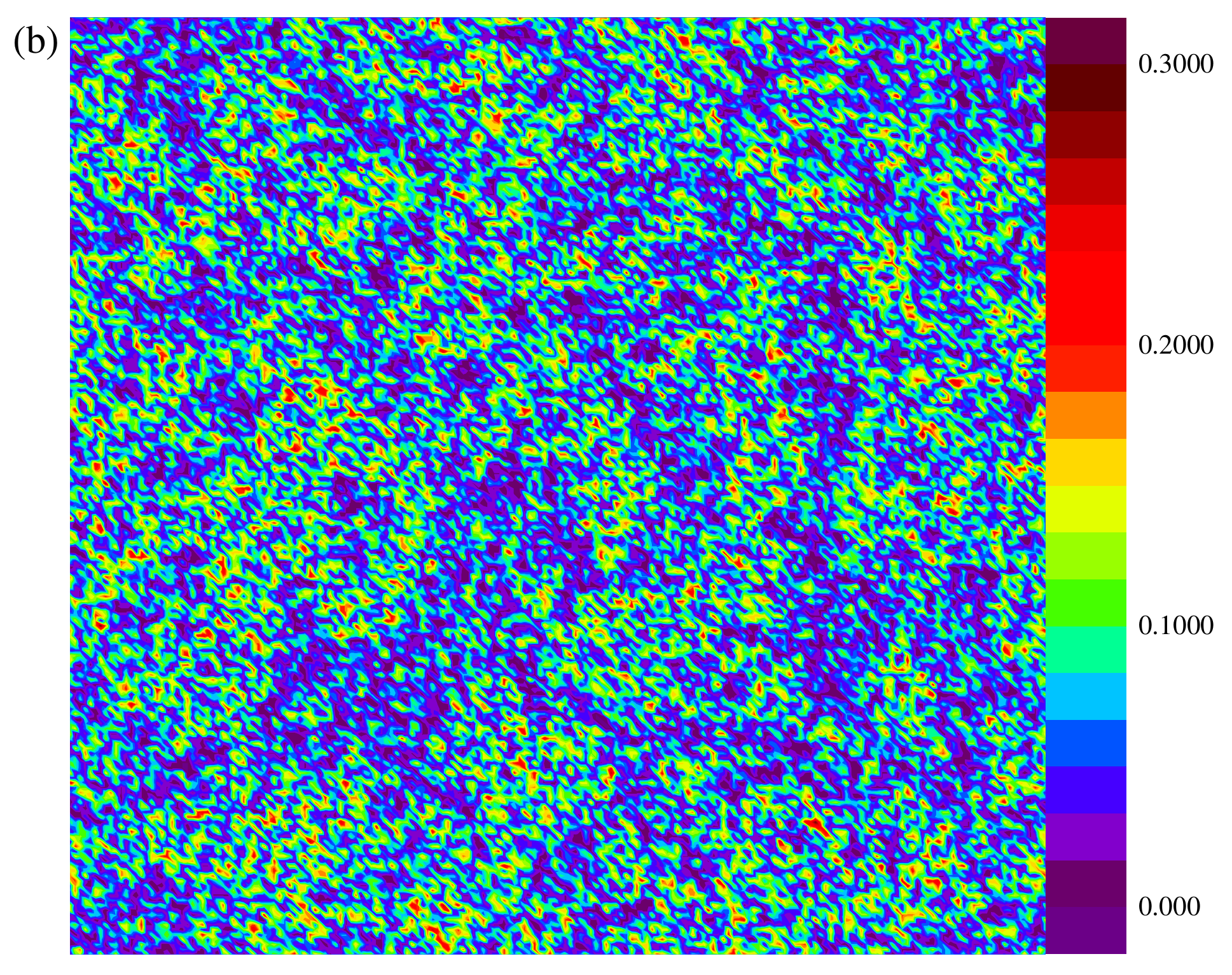}\\
	\includegraphics[width=8 cm]{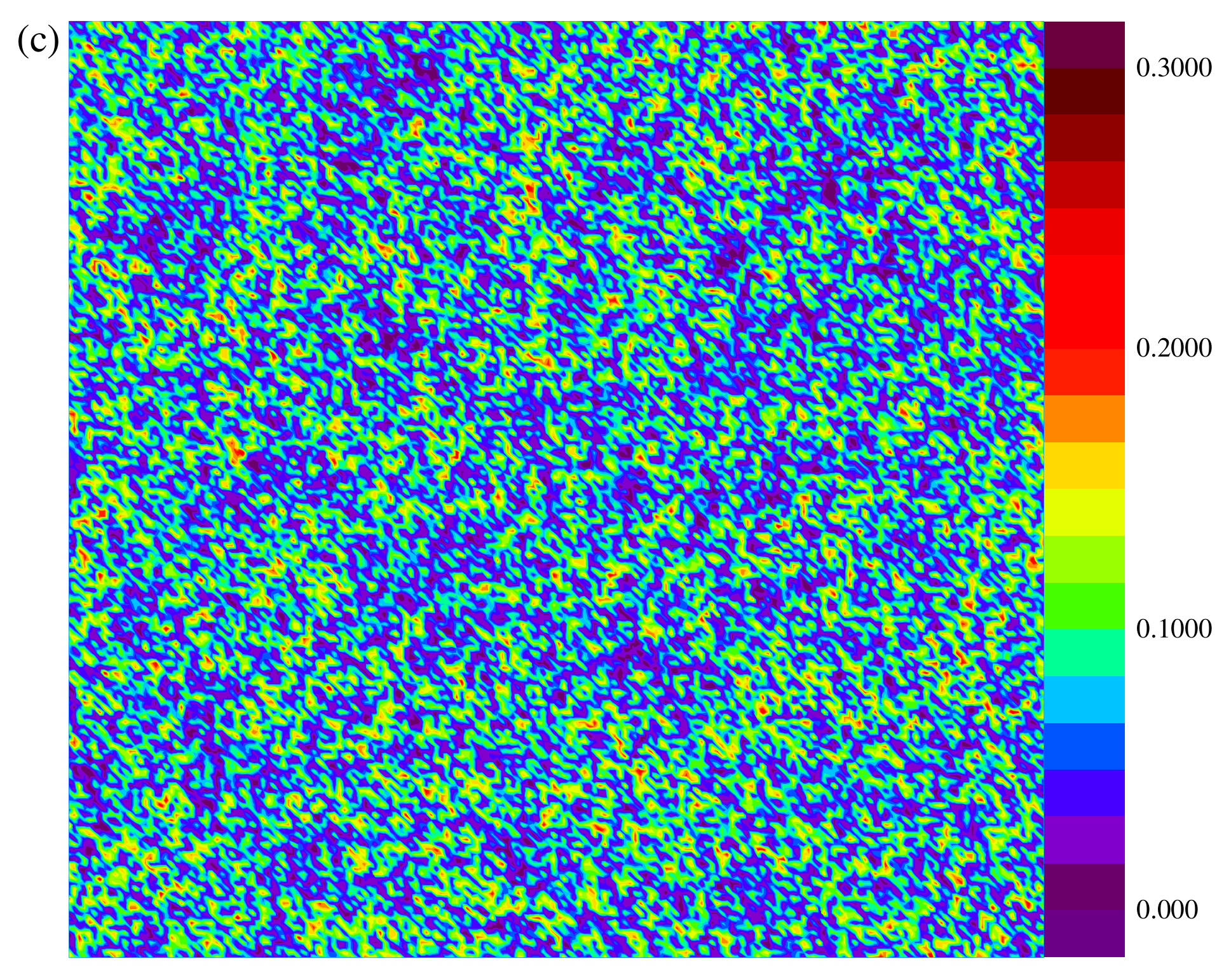}
	\caption{\label{fig:configurations_new} In full analogy with Fig.~\ref{fig:configurations} we show snapshots of the period-averaged quadrupole moment conjugate to the crystal-field coupling $\Delta$. Simulation parameters are exactly the same as those used in Figs. \ref{fig:configurations}(a)--\ref{fig:configurations}(c).}
\end{figure}

\begin{figure}[t]
	\centering
	\includegraphics[width=8 cm]{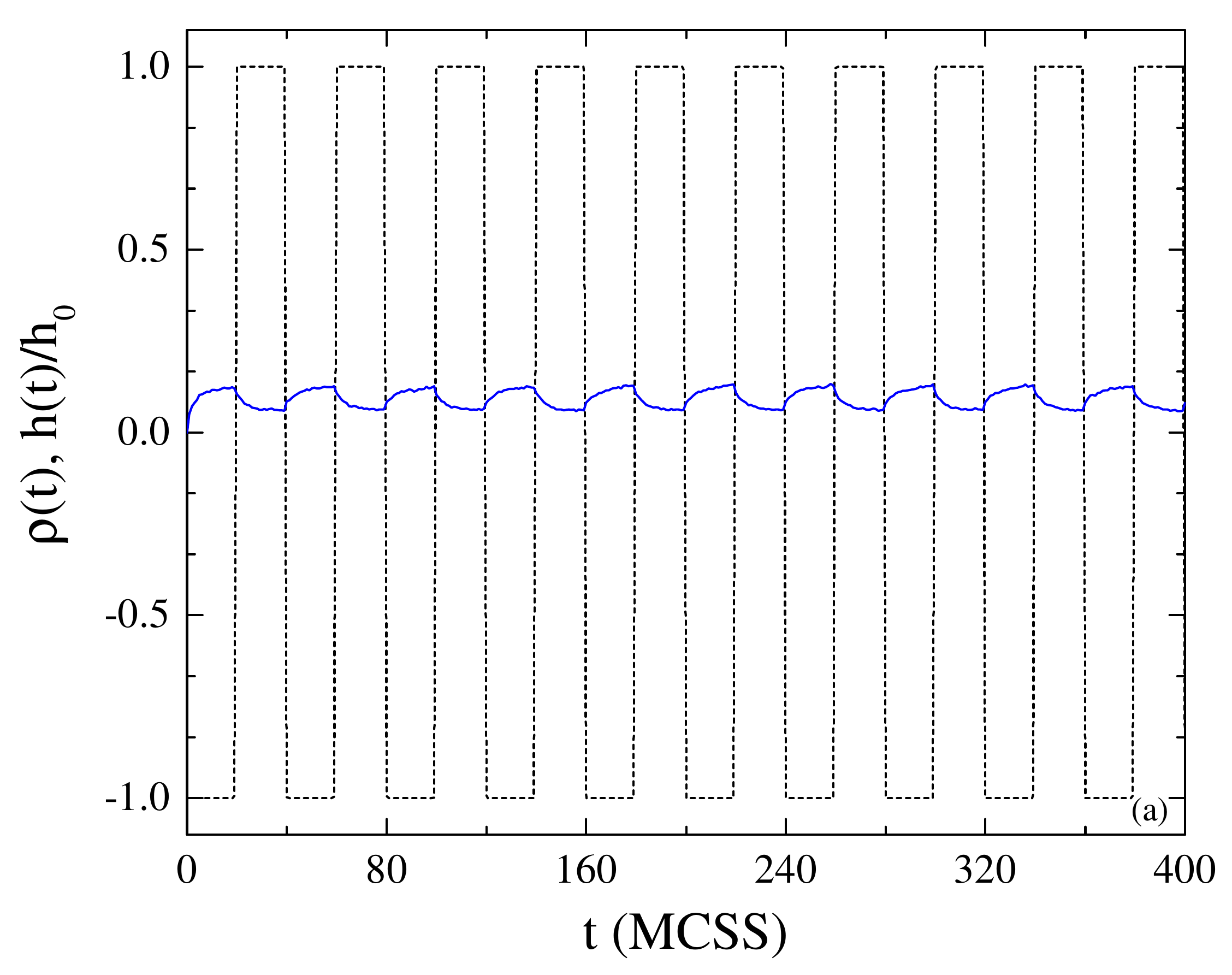}\\
	\includegraphics[width=8 cm]{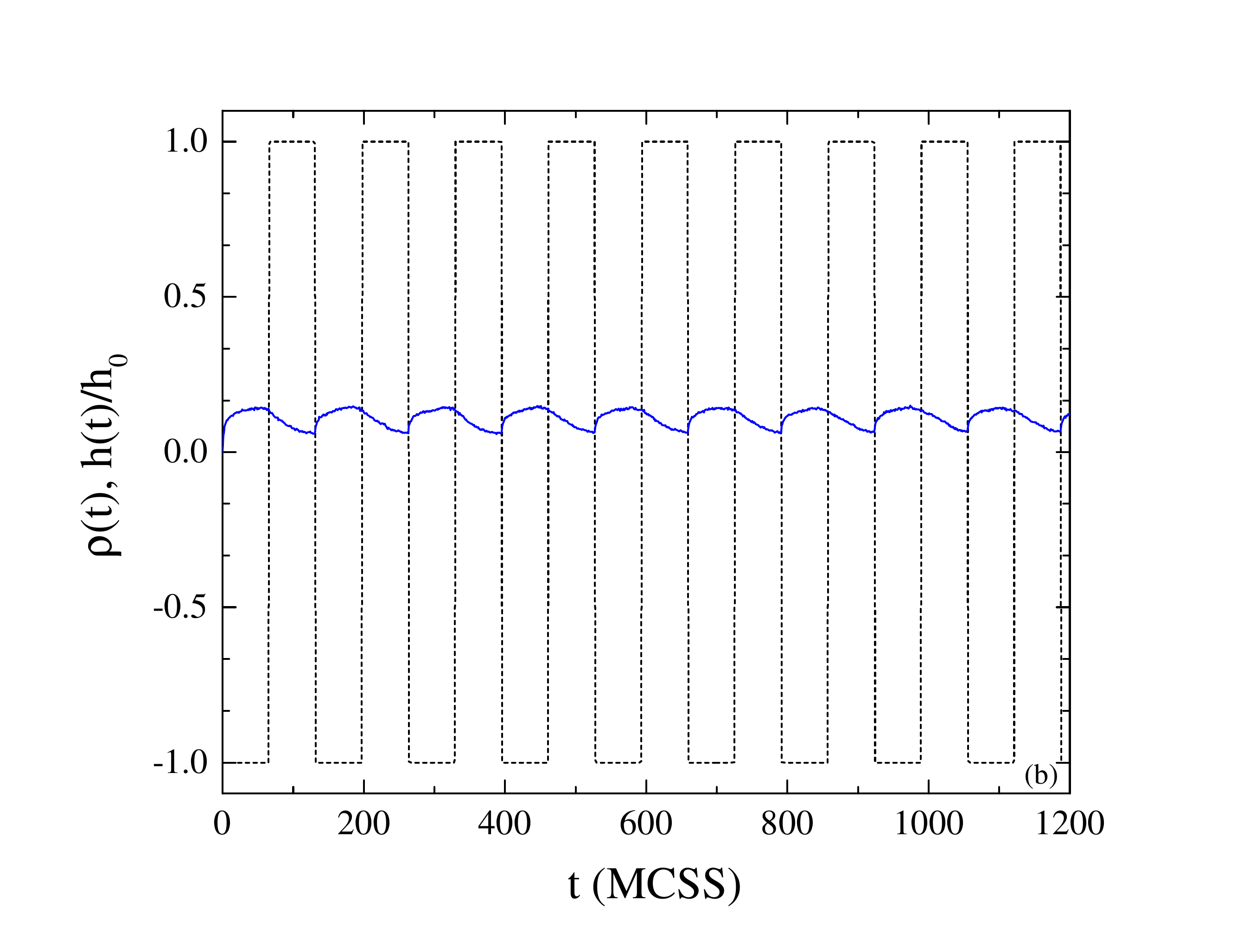}\\
	\includegraphics[width=8 cm]{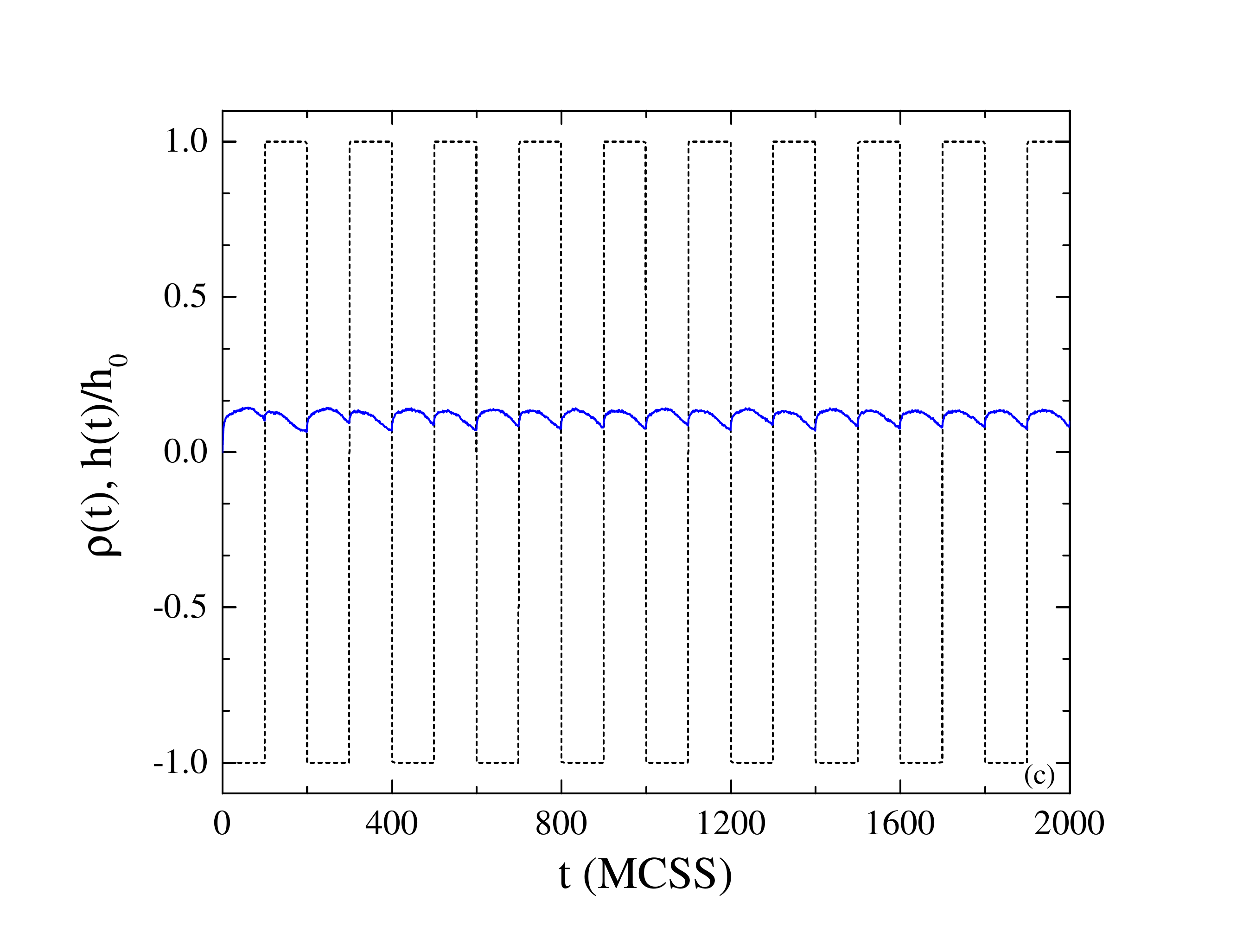}
	\caption{\label{fig:timeseries_new} Time series of the
		order parameter conjugate to the crystal field $\rho(t)$ (solid blue curves) of the kinetic random $p=1/2$ Blume-Capel model under the presence of a square-wave magnetic field (black
		dashed lines) for $L = 192$ at $\Delta = 1$, for three values of the half period
		as in Fig.~\ref{fig:timeseries}: (a) $t_{1/2} = 20$ MCSS, (b) $t_{1/2} = 66$ MCSS, and (c) $t_{1/2} = 100$ MCSS. Again for the
		sake of clarity the ratio $h(t)/h_{0}$ is displayed.}
\end{figure}

\begin{figure}
	\includegraphics[width=11 cm]{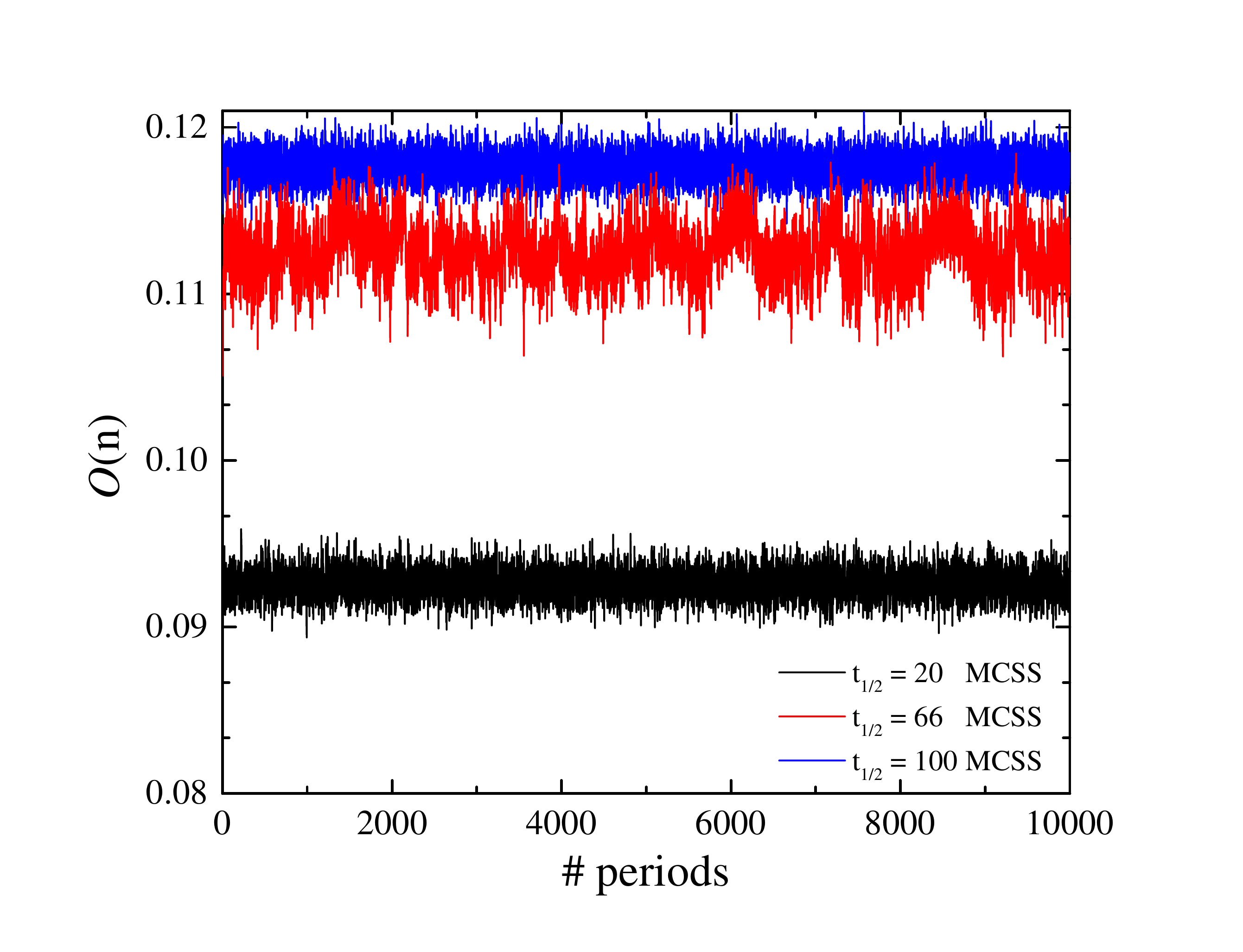}
	\caption{Period dependencies of the
		dynamic quadrupole moment $O$ of the kinetic random $p=1/2$ Blume-Capel model for
		$L = 192$ at $\Delta = 1$. Results are shown for the three characteristic cases of
		the half period of the external field, following
		Fig.~\ref{fig:timeseries_new}.
		\label{fig:period_new}}
\end{figure}

\begin{figure}[t]
	\centering
	\includegraphics[width=8 cm]{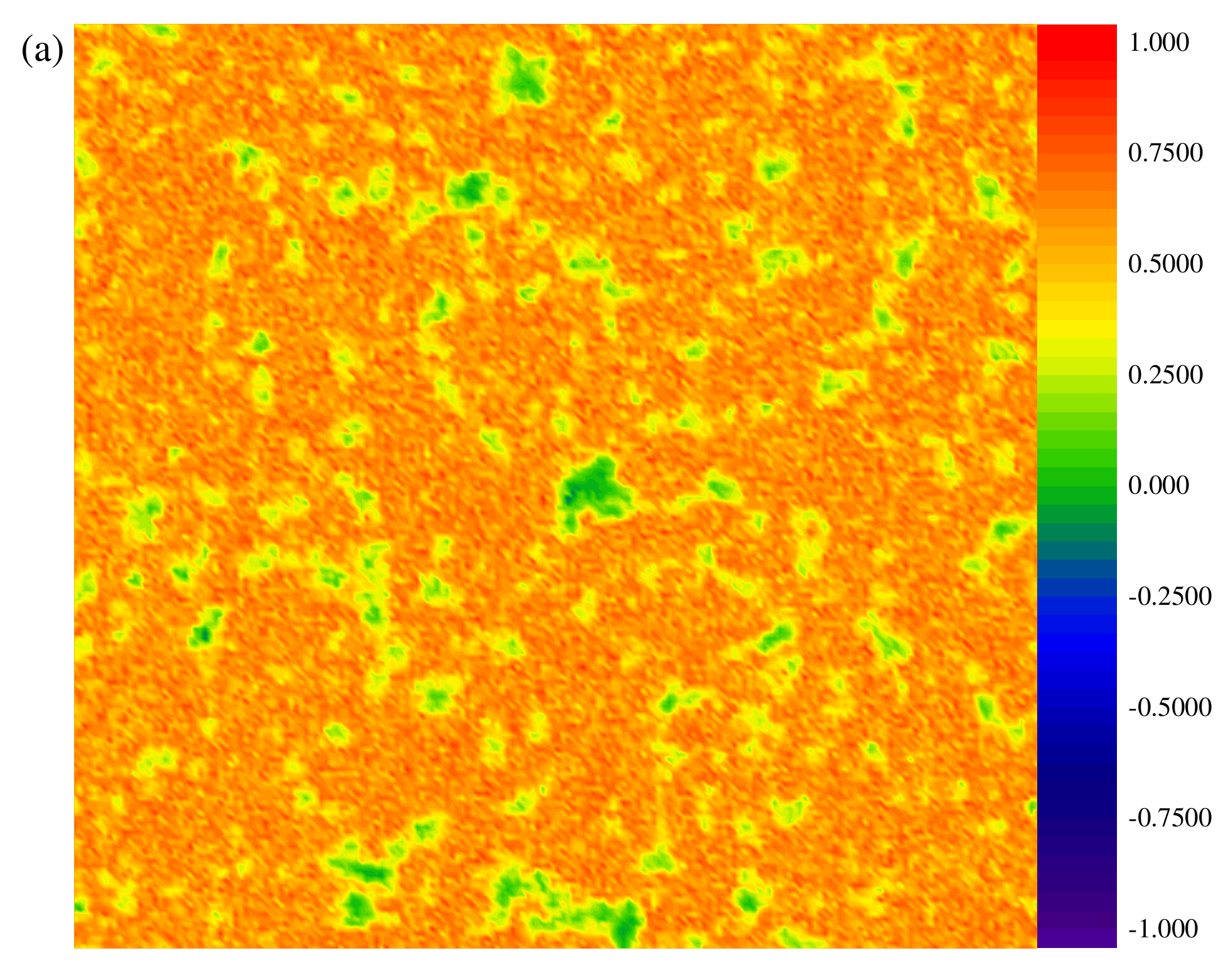}\\
	\includegraphics[width=8 cm]{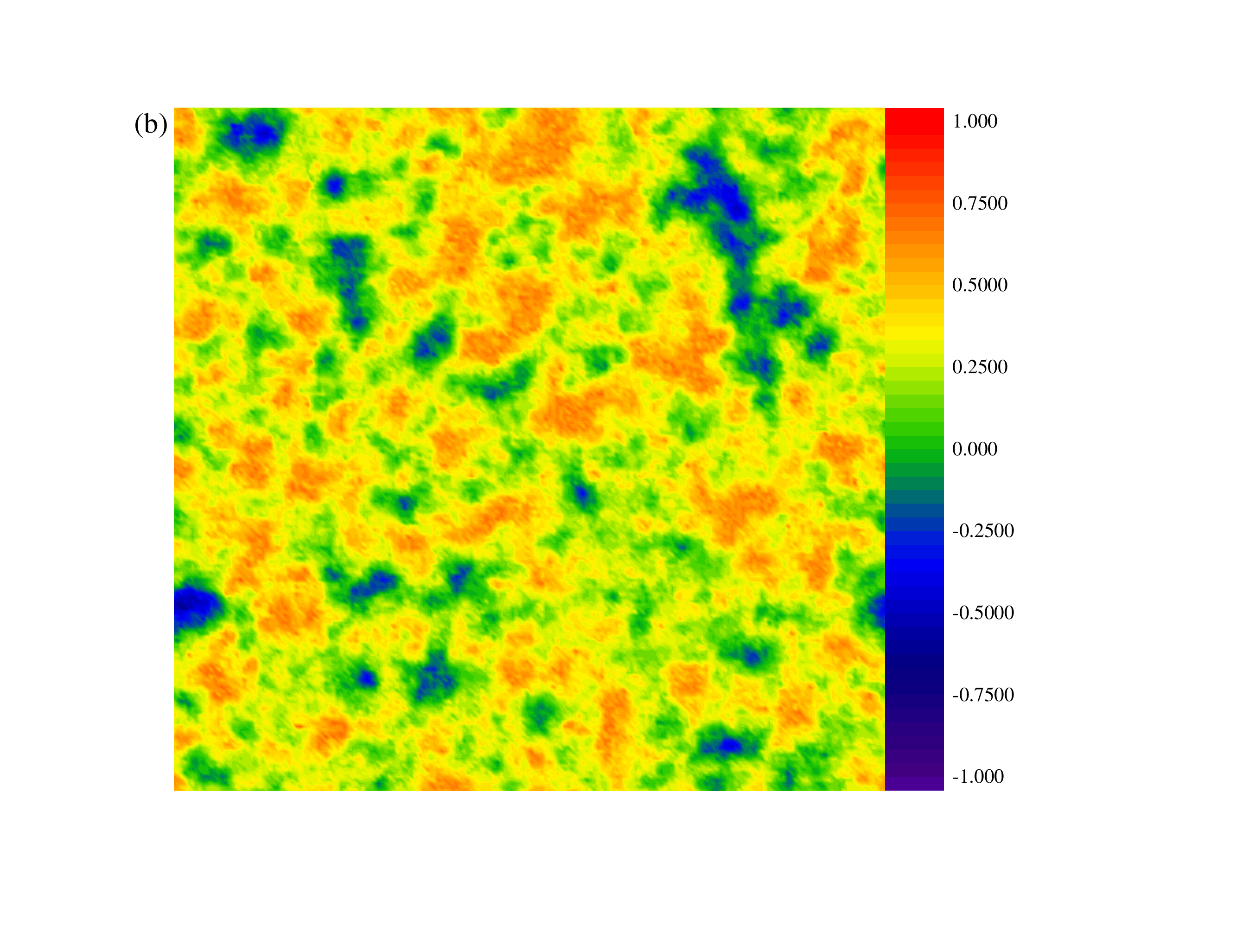}\\
	\includegraphics[width=8 cm]{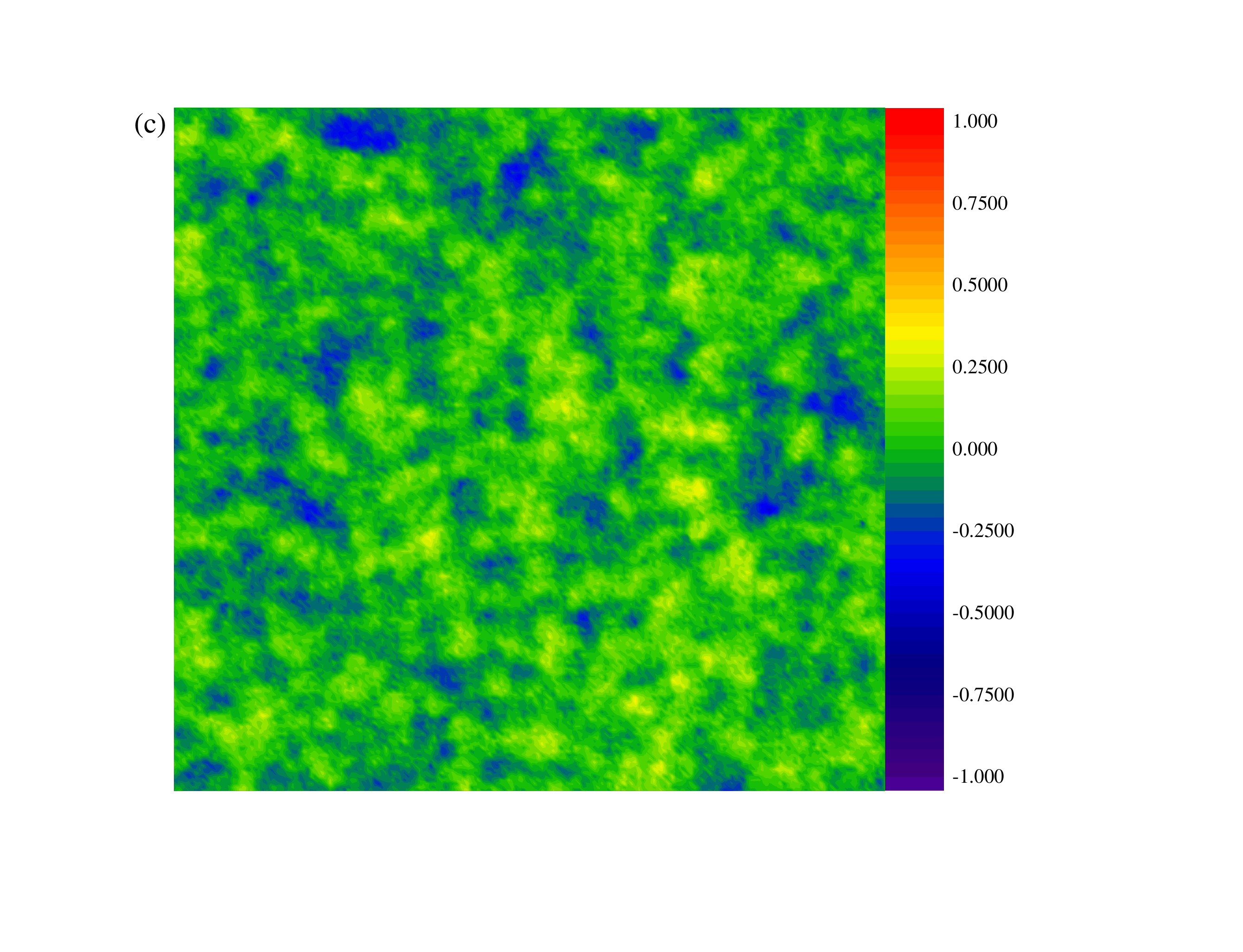}
	\caption{\label{fig:configurations_low_p} Configurations of the local
		dynamic order parameter $\{Q_{x}\}$ of the kinetic random Blume-Capel
		model for $L = 192$, $p=0.02$, and $\Delta = 2$. Note that for this set of ($p$, $\Delta$)--parameters we approximated the critical half period of the system to be $t_{1/2}^{\rm c} \approx 53$, from the peak positions of the corresponding dynamic susceptibility and heat-capacity curves. Three panels are shown: (a) $t_{1/2} = 20$ MCSS $ < t_{1/2}^{\rm c}$ -- dynamically ordered phase, (b)
		$t_{1/2} = 53$ MCSS $\approx t_{1/2}^{\rm c}$ -- near the dynamic
		phase transition, and (c) $t_{1/2} = 100$ MCSS $ > t_{1/2}^{\rm
			c}$ -- dynamically disordered phase. }
\end{figure}

\begin{figure}[t]
	\centering
	\includegraphics[width=8 cm]{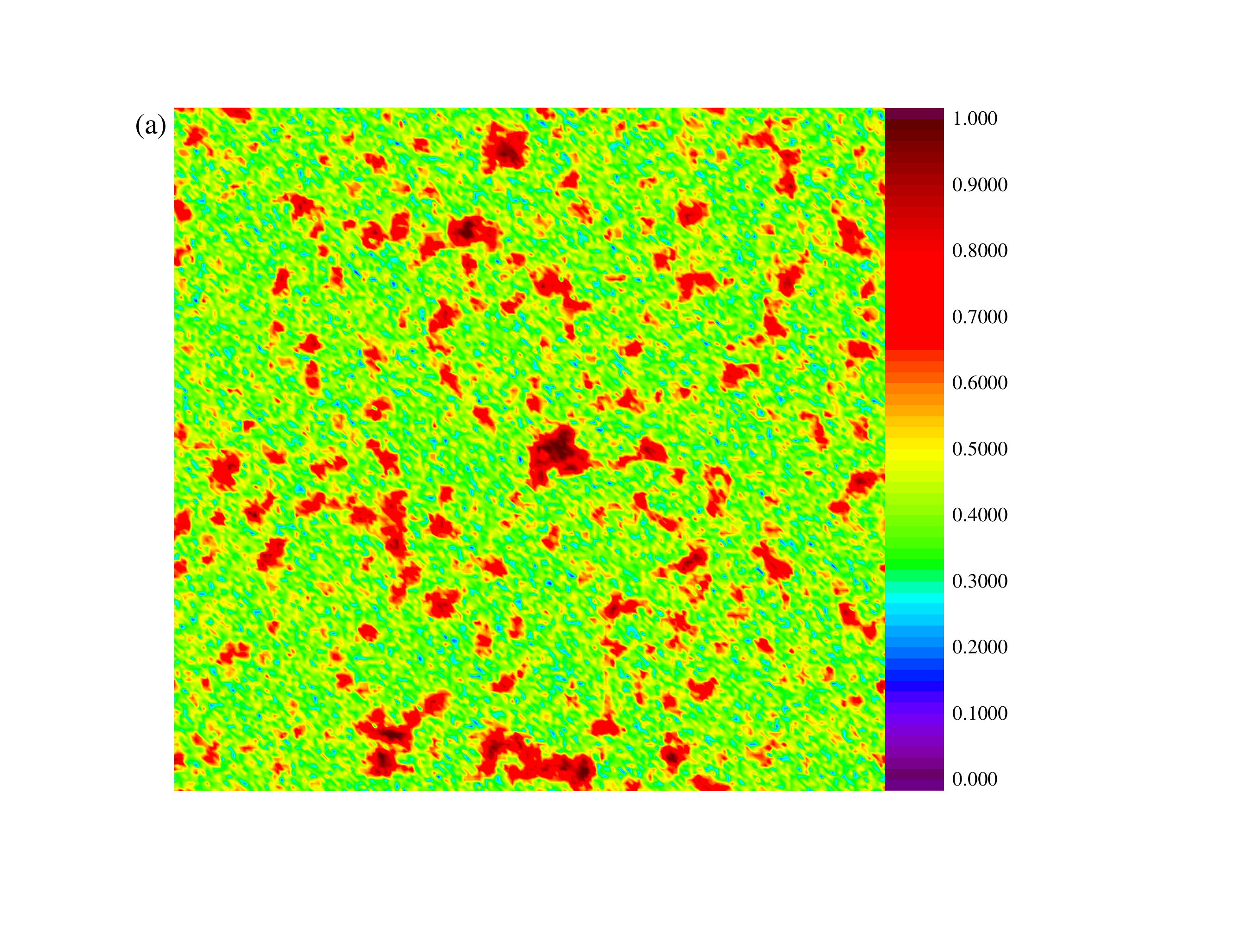}\\
	\includegraphics[width=8 cm]{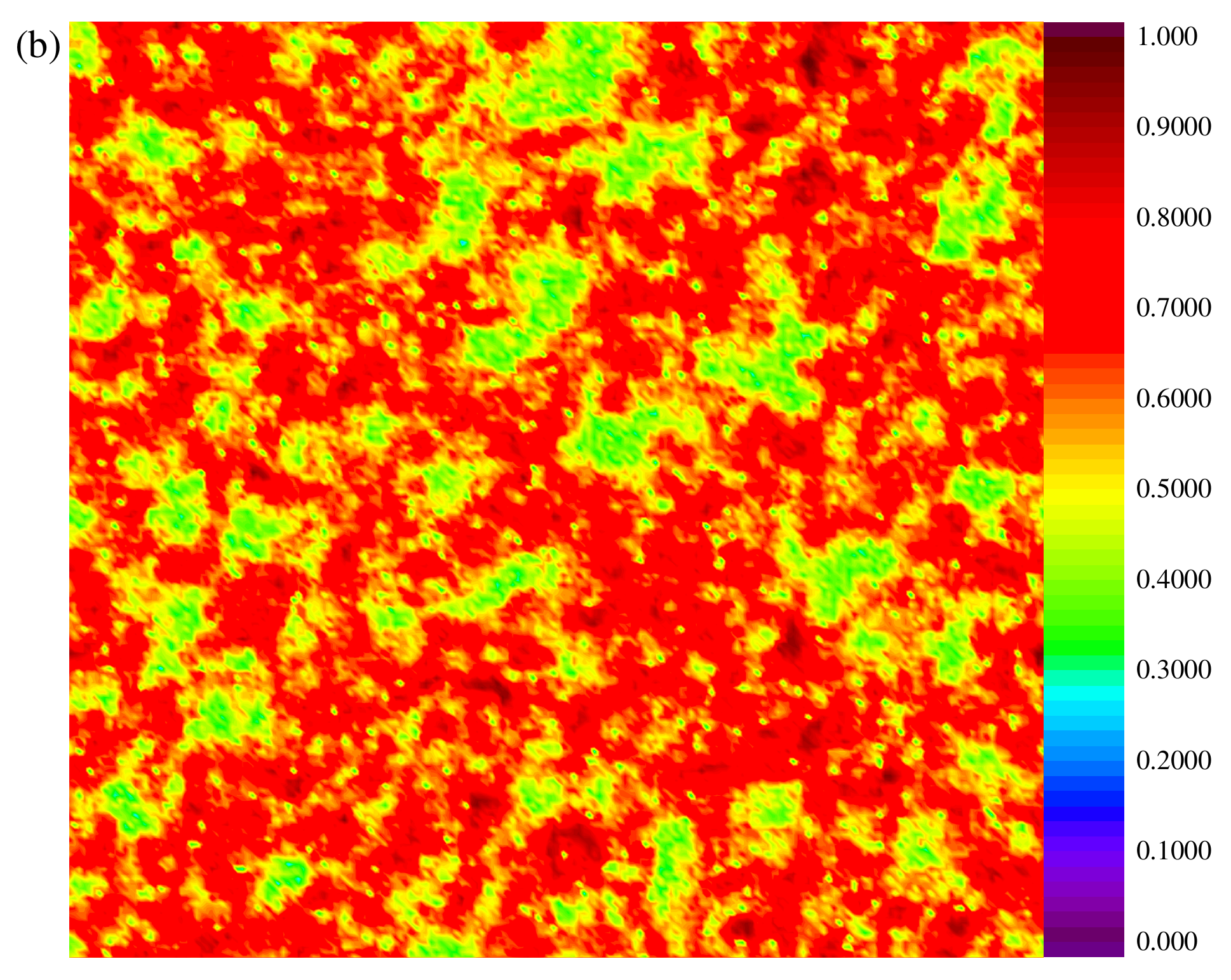}\\
	\includegraphics[width=8 cm]{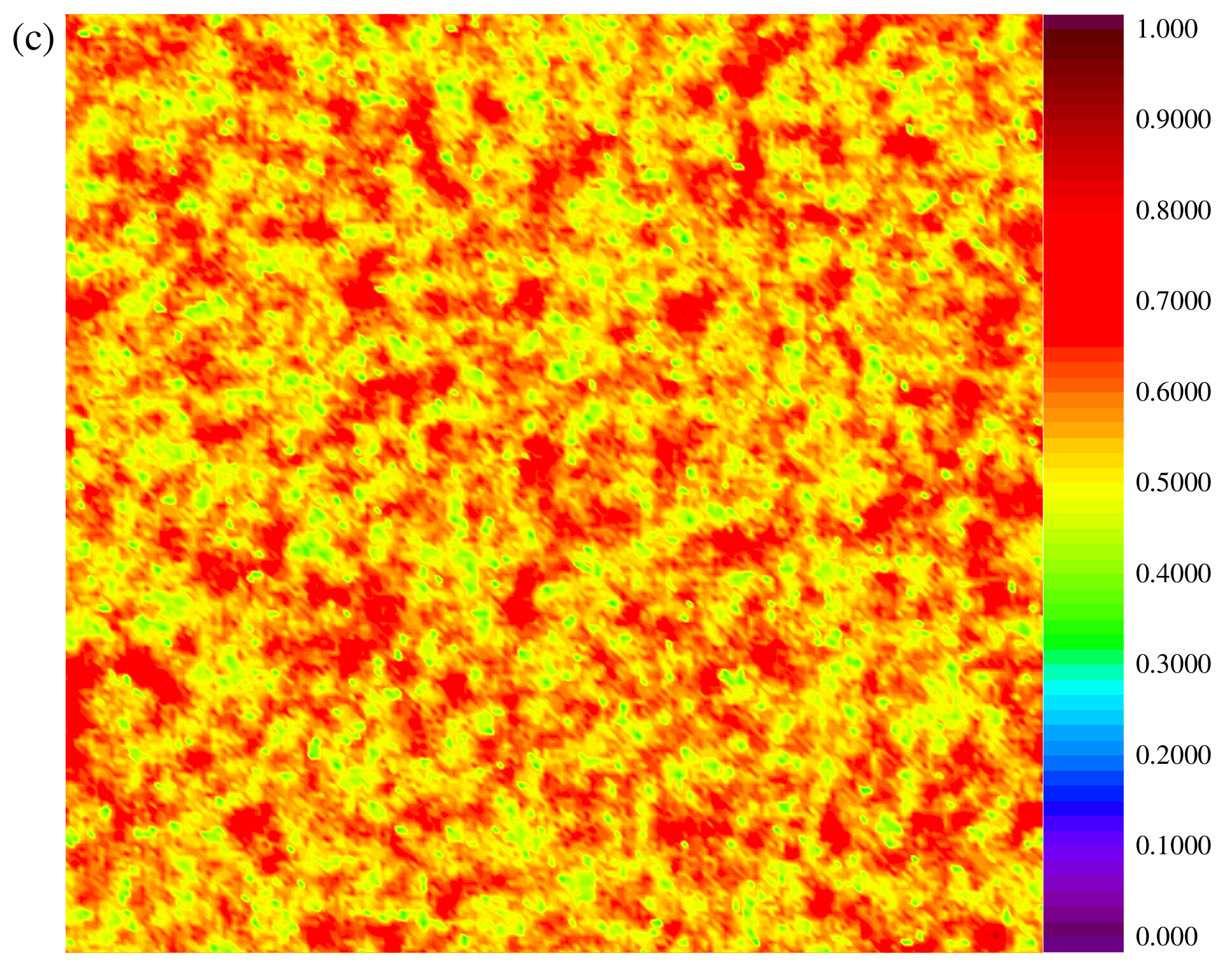}
	\caption{\label{fig:configurations_low_p_new} In full analogy with Fig.~\ref{fig:configurations_low_p} we show snapshots of the period-averaged quadrupole moment conjugate to the crystal-field coupling $\Delta$. Simulation parameters are exactly the same as those used in Figs. \ref{fig:configurations_low_p}(a)--\ref{fig:configurations_low_p}(c). }
\end{figure}


\begin{figure}
	\includegraphics[width=11 cm]{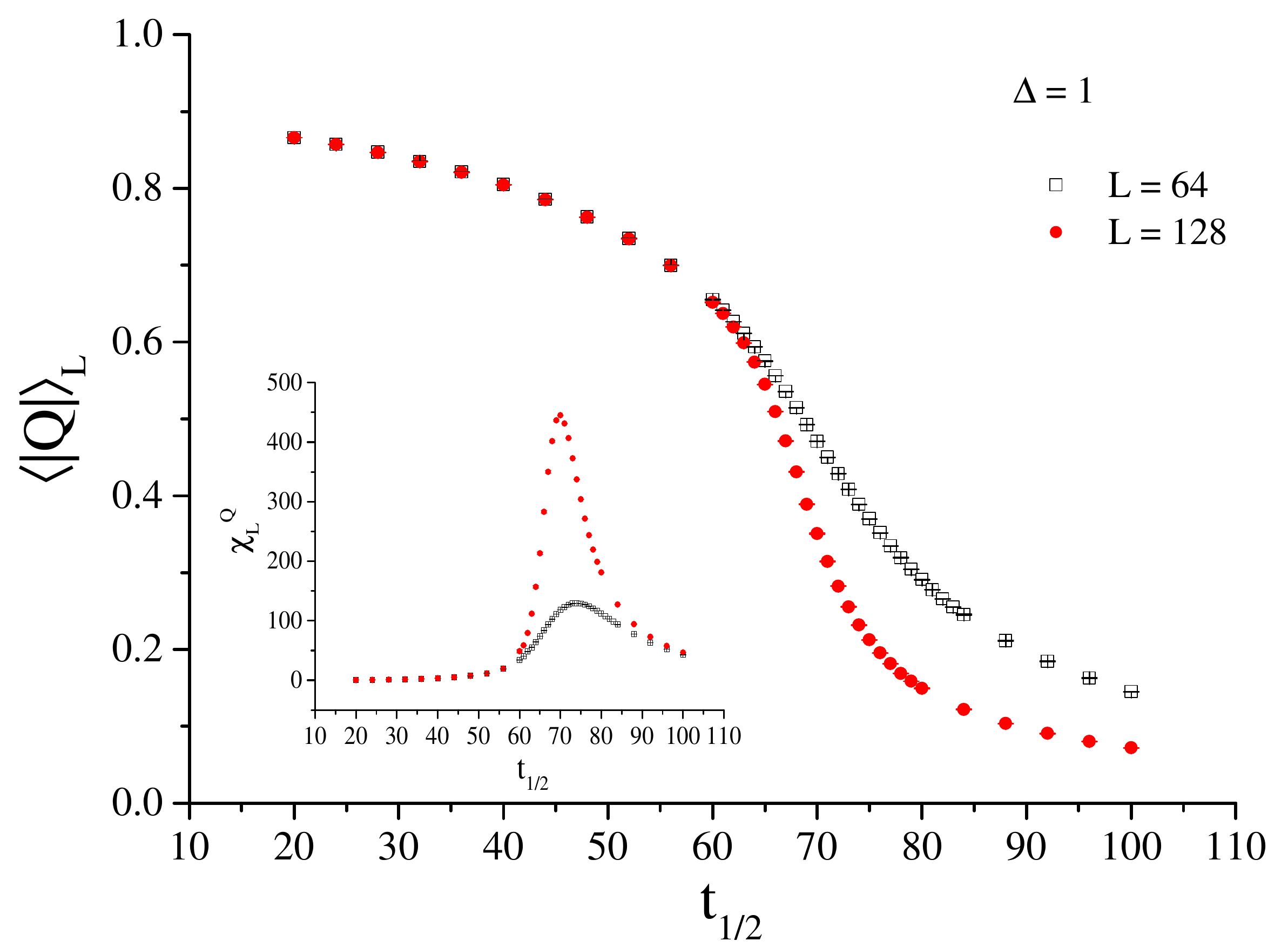}
	\caption{Typical dynamic order parameter $\langle |Q| \rangle_{L}$ (main panel) and susceptibility $\chi_{L}^{Q}$ (inset) curves of the kinetic random $p=1/2$ Blume-Capel model at $\Delta = 1$ and for two systems with linear sizes $L = 64$ (open black squares) and $L = 128$ (filled red circles).
		\label{fig:curves_magnetic}}
\end{figure}

\begin{figure}
	\includegraphics[width=11 cm]{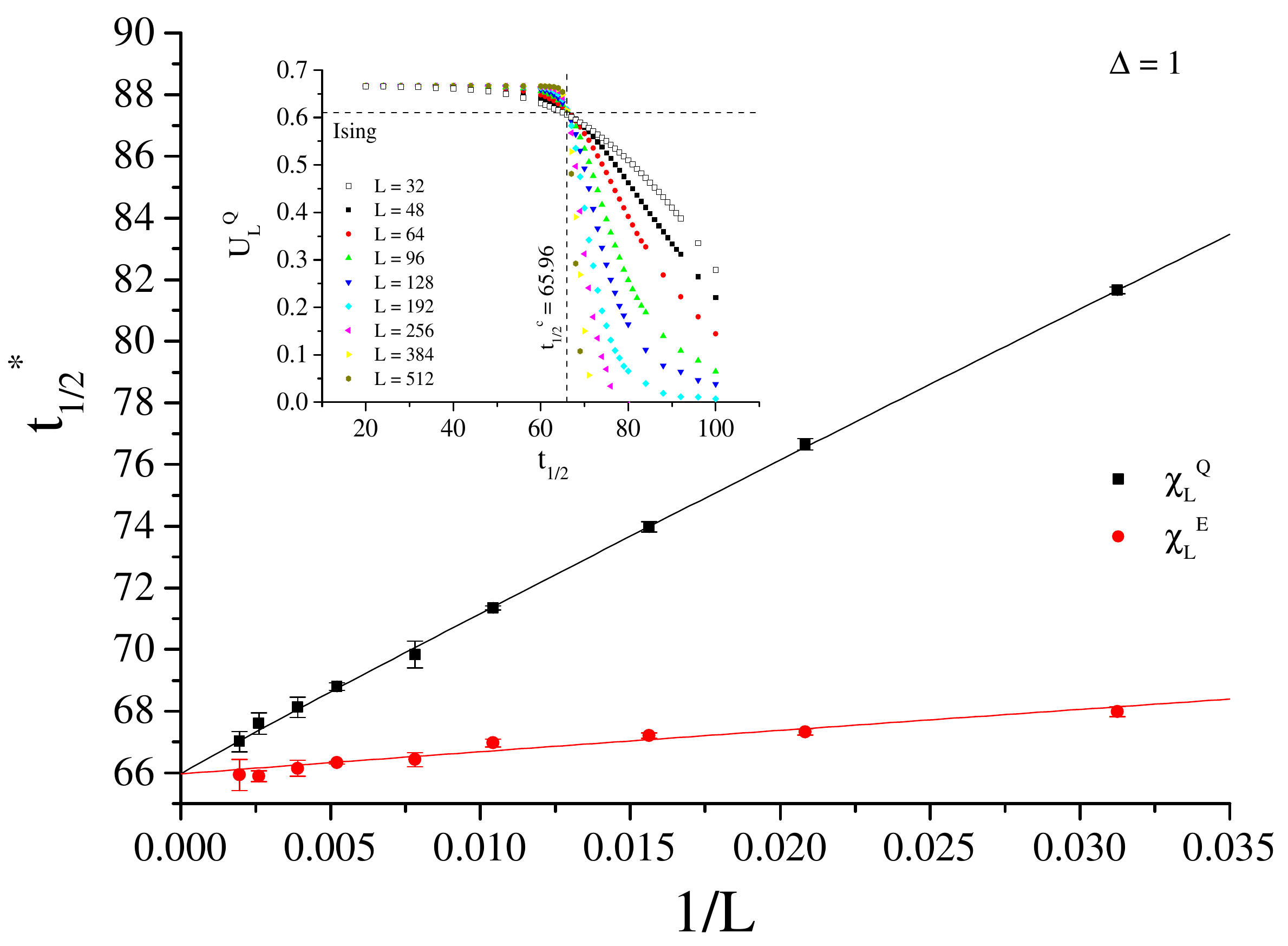}
	\caption{Shift behavior of the two pseudocritical half periods $t_{1/2}^{\ast}$ corresponding to the maxima of the dynamic susceptibility (filled black squares) and heat capacity (filled red circles) of the kinetic random $p=1/2$ Blume-Capel model at $\Delta = 1$. The inset illustrates the half-period dependency of the corresponding fourth-order Binder cumulant $U_{L}^{Q}$.
		\label{fig:pseudocritical}}
\end{figure}

\begin{table}
	\caption{A summary of critical parameters describing the dynamic phase transition of the square-lattice kinetic Blume-Capel model in a quenched random crystal field. Note that the values of $\Delta$ considered in the current work, given the randomness distribution~(\ref{eq:bimodal}) with $p=1/2$, correspond to the second-order transition regime of the model's equilibrium phase diagram. One needs very small values of $p$, \emph{i.e.}, $p\leq 0.1$, in order to reach the originally first-order transition regime at high values of $\Delta \approx 2$, see Figs. 9 - 13 in Ref.~\cite{Vatansever20}.}\label{table}
	\begin{center}
		\begin{tabular}{ |c|c|c|c| }
			\hline
			$\Delta$ & $t_{1/2}^{\rm c}$ & $\nu$ & $\gamma/\nu$ \\
			\hline
			$0.5$ & $72.41(9)$ & $1.00(3)$ & $1.75(1)$\\
			\hline
			$1$ & $65.96(6)$ & 1.03(3) & $1.76(1)$ \\
			\hline
			$2$ & $47.61(7)$ & $1.05(7)$ & $1.75(2)$ \\
			\hline
		\end{tabular}
	\end{center}
\end{table}

\begin{figure}
	\includegraphics[width=11 cm]{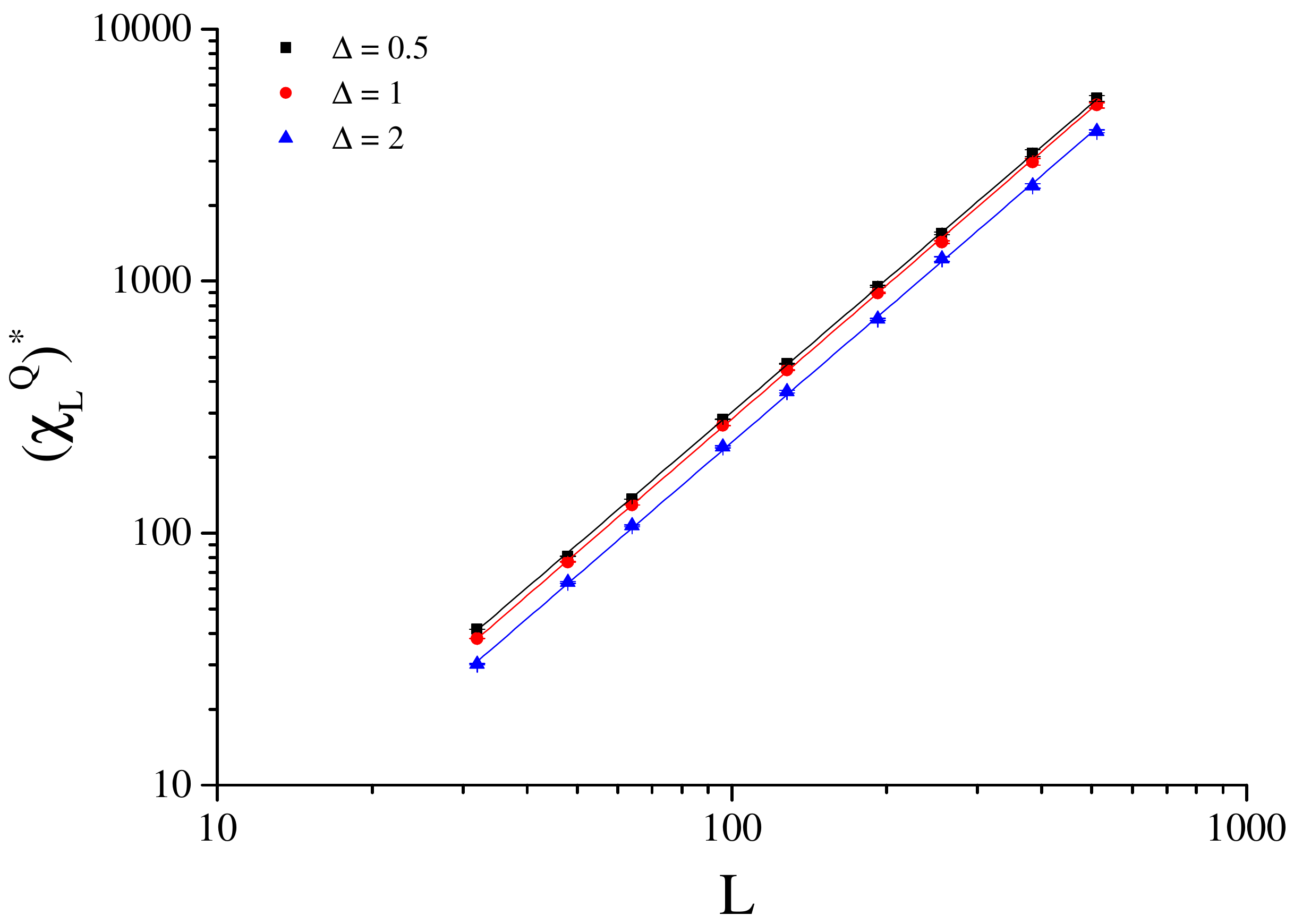}
	\caption{Finite-size scaling behavior of the dynamic susceptibility
		maxima $(\chi_{L}^Q)^{\ast}$ of the kinetic random $p=1/2$ Blume-Capel model. Results for three values of $\Delta$ are shown in a log-log scale.
		\label{fig:chi}}
\end{figure}

\begin{figure}
	\includegraphics[width=11 cm]{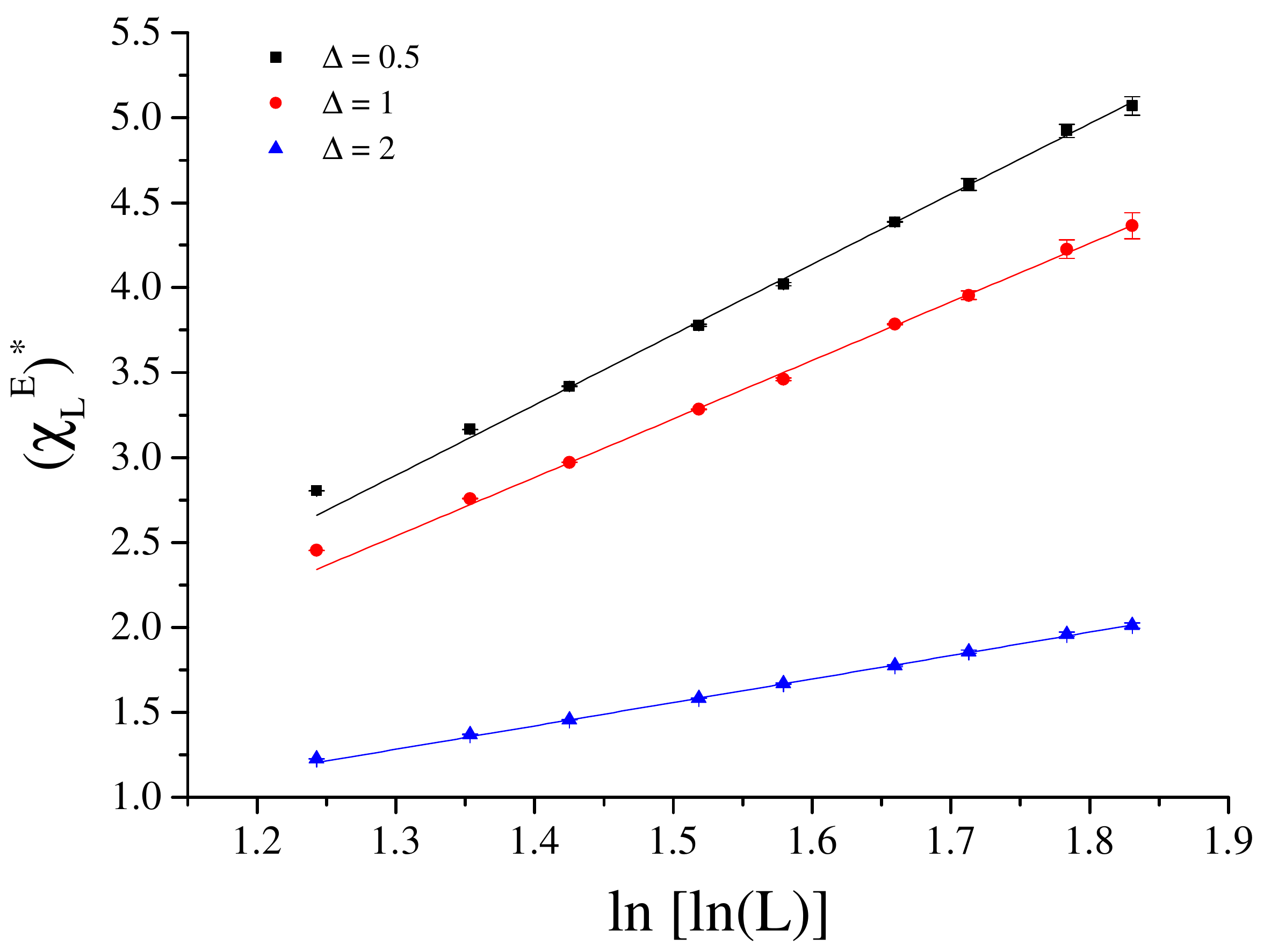}
	\caption{Double logarithmic scaling behavior of the heat-capacity maxima $(\chi_{L}^E)^{\ast}$ of the kinetic random $p=1/2$ Blume-Capel model for three values of $\Delta$, as indicated in the panel.
		\label{fig:sp_heat}}
\end{figure}

\end{document}